\documentclass{aa}
\usepackage{ amsmath }
\usepackage{ amssymb }
\usepackage{graphicx}
\usepackage{aas_macros}
\usepackage{caption}
\usepackage{subcaption}
\usepackage{morefloats}
\usepackage[varg]{txfonts}
\usepackage{dblfloatfix}
\usepackage{fixltx2e}
\bibpunct{(}{)}{;}{a}{}{,}

\begin{document}
\title{Mid-{\it J} CO shock tracing observations of infrared dark clouds I \thanks{{\it Herschel} is an ESA space observatory with science instruments provided by European-led Principal Investigator consortia and with important participation from NASA.}}

\author{A. Pon\inst{\ref{MPE}}
\and P. Caselli\inst{\ref{MPE}} 
\and D. Johnstone\inst{\ref{JAC},\ref{HIA},\ref{UVic}}
\and M. Kaufman\inst{\ref{SanJose},\ref{Ames}}
\and M. J. Butler\inst{\ref{Zurich}}
\and F. Fontani\inst{\ref{INAF}}
\and I. Jim\'{e}nez-Serra\inst{\ref{ESO}} 
\and J. C. Tan\inst{\ref{Florida}}}

\institute{Max-Planck-Institut f\"{u}r extraterrestrische Physik, Giessenbachstrasse 1, 85748 Garching, Germany \email{andyrpon@mpe.mpg.de}\label{MPE}
\and Joint Astronomy Centre, 660 North A'ohoku Place, University Park, Hilo, HI 96720, USA\label{JAC}
\and NRC Herzberg Astronomy and Astrophysics, 5071 West Saanich Road, Victoria, BC V9E 2E7, Canada\label{HIA}
\and Department of Physics and Astronomy, University of Victoria, PO Box 3055 STN CSC, Victoria, BC V8W 3P6, Canada\label{UVic}
\and Department of Physics and Astronomy, San Jose State University, One Washington Square, San Jose, CA 95192-0106, USA\label{SanJose}
\and Space Science and Astrobiology Division, MS 245-3, NASA Ames Research Center, Moffett Field, CA 94035, USA\label{Ames}
\and Institute of Theoretical Physics, University of Z\"{u}rich, 8057 Z\"{u}rich, Switzerland\label{Zurich}
\and INAF- Osservatorio Astrofisico di Arcetri, Largo E. Fermi 5, Firenze 50125, Italy\label{INAF}
\and European Southern Observatory, Karl-Schwarzschild-Str. 2, 85748, Garching bei Muenchen, Germany\label{ESO}
\and Departments of Astronomy \& Physics, University of Florida, Gainesville, FL 32611, USA\label{Florida}}

\abstract{Infrared dark clouds (IRDCs) are dense, molecular structures in the interstellar medium that can harbour sites of high-mass star formation. IRDCs contain supersonic turbulence, which is expected to generate shocks that locally heat pockets of gas within the clouds. We present observations of the CO $J$ = 8-7, 9-8, and 10-9 transitions, taken with the {\it Herschel} Space Observatory, towards four dense, starless clumps within IRDCs (C1 in G028.37+00.07, F1 and F2 in G034.43+0007, and G2 in G034.77-0.55). We detect the CO $J$ = 8-7 and 9-8 transitions towards three of the clumps (C1, F1, and F2) at intensity levels greater than expected from photodissociation region (PDR) models. The average ratio of the 8-7 to 9-8 lines is also found to be between 1.6 and 2.6 in the three clumps with detections, significantly smaller than expected from PDR models. These low line ratios and large line intensities strongly suggest that the C1, F1, and F2 clumps contain a hot gas component not accounted for by standard PDR models. Such a hot gas component could be generated by turbulence dissipating in low velocity shocks.}


\maketitle

\section{INTRODUCTION}
\label{introduction}

Giant molecular clouds contain significant supersonic turbulent motions and this turbulence may play a key role in supporting molecular clouds against collapse on large scales, as well as in generating small-scale density perturbations that can then locally collapse to form stars (e.g., \citealt{Padoan02, McKee07}). Numerical simulations of magnetohydrodynamic (MHD) turbulence, similar to that observed in molecular clouds, show that such turbulence decays on the order of the crossing time at the driving scale \citep{Gammie96, MacLow98, Stone98, MacLow99, Padoan99, Ostriker01}. In typical Galactic star-forming regions, the volume averaged turbulent dissipation rate is on the order of the cosmic ray heating rate \citep{Pon12Kaufman}. This turbulent heating, however, only occurs within localized shock fronts, such that these shocks create a hot gas component with a low volume filling factor, thereby producing a very different spectral signature than would be expected from uniformly heated gas \citep{Pon12Kaufman}. 

\citet{Pon12Kaufman} present MHD, C-type shock models for 2 and 3 km s$^{-1}$ shocks propagating into gas with densities between 10$^{2.5}$ and 10$^{3.5}$ cm$^{-3}$. By scaling these models to the expected turbulent energy dissipation rate of a molecular cloud (e.g., \citealt{Basu01}), \citet{Pon12Kaufman} predict the integrated intensities of shock excited lines from low-mass star-forming regions. They show that most of the energy in these shocks is radiated by rotational transitions of CO and, by comparing their shock results to the photodissociation region (PDR) models of \citet{Kaufman99}, they predict that mid-$J$ CO lines (J$_{\mbox{upper}} \ge 6$) should be dominated by emission from shocked gas, despite this hot shocked gas only having volume filling factors of the order of 0.1\%. Such excess emission in the CO $J$ = 6 $\rightarrow$ 5 line, consistent with an extra hot gas component, has already been detected in the Perseus B1-East 5 low-mass star-forming region \citep{Pon14Kaufman} and in the Taurus molecular cloud \citep{Larson15}. 

Infrared dark clouds (IRDCs) are molecular gas structures, which are usually filamentary, that appear as dark lanes against the bright mid-infrared ($\sim 8 \mu$m) Galactic background due to their high column densities, with visual extinctions approaching 100 mag in some regions (e.g., \citealt{Rathborne06, Butler12}). Some IRDCs are associated with high-mass star-forming regions. Four particularly dense, infrared dark clumps within IRDCs (C1 in G028.37+00.07, F1 and F2 in G034.43+00.24, and G2 in G034.77-00.55) were selected from the \citet{Butler12} sample as prime candidates for evolved clumps without any embedded young stellar objects. These four clumps are ideal candidates to search for a small volume filling factor, hot gas component caused by the dissipation of turbulence, since they are unlikely to contain significant internal heating sources that can mimic the heating caused by low velocity shocks and because they have large columns of gas in which the shocks can exist.  

The four clumps observed are described in more detail in Sect.\ \ref{sources}. In Sect.\ \ref{Herschel}, observations taken with the {\it Herschel Space Observatory} of the $^{12}$CO $J$ = 8 $\rightarrow$ 7, 9 $\rightarrow$ 8, and 10 $\rightarrow$ 9 transitions towards the C1, F1, F2, and G2 quiescent clumps are presented. Photodissociation region (PDR) models are compared to the observations in Sect.\ \ref{PDR}. The implications of such a comparison are discussed in Sect.\ \ref{discussion} and we summarize our main findings in Sect.\ \ref{conclusions}. 

\section{SOURCES}
\label{sources}

We consider clumps to be structures that are large enough to fragment into star clusters and cores to be structures likely to form a single star. To match such a theoretically based classification system to observed structures in IRDCs, we choose to refer to objects on the size scale of a parsec as clumps, and objects on the size scale of 0.1 pc as cores. Such a naming convention helps us to clarify the size scale of the object being described and to differentiate between substructures within regions. With this working, observational definition, clumps will typically have masses on the order of hundreds of solar masses \citep{Butler09, Butler12} while cores will have masses on the order of tens of solar masses \citep{Tan13, Butler14}. 

The four clumps examined in this paper are the clumps studied by \citet{Tan13}. The locations of these four clumps, as given by \citet{Butler12}, are listed in Table \ref{table:targets} and the clumps will be referred to as C1, F1, F2, and G2 \citep{Butler09, Butler12}. These clumps are spread between three different IRDCs, the C (G028.37+00.07), F (G034.43+00.24), and G (G034.77-0.55) IRDCs of \citet{Butler09}. IRDC C is also known as the Dragon Nebula \citep{Wang15Thesis}. These clumps were selected by \citet{Tan13} for five reasons. First, the clumps are well studied, having been surveyed in continuum emission \citep{Rathborne06, Chambers09}, mid-infrared extinction \citep{Butler09, Butler12}, and molecular line emission, including, but not limited to, N$_2$H$^+$ and N$_2$D$^+$ emission \citep{Fontani11}. Second, the clumps are dark at 24 and 70 microns in {\it Spitzer} MIPSGAL images \citep{Carey09}, indicating the lack of massive embedded protostars.  Third, the clumps are in relatively quiescent environments, indicated by the lack of mid-infrared sources within 20 arcsec \citep{Benjamin03, Tan13}. Fourth, the clumps have large mass surface densities \citep{Butler09,Butler12} and fifth, they have large deuterium fractions measured by the N$_2$D$^+$ to N$_2$H$^+$ column density ratio \citep{Fontani11}. These four clumps are thus prime candidates to be sites of star formation in the near future, but are unlikely to currently contain massive protostars. 

\begin{table*}
\begin{minipage}{\textwidth}
\caption{Target Locations}
\begin{center}
\begin{tabular}{cccccccc}
\hline
\hline
Clump & Cloud & RA (J2000) & Dec(J2000) & $l$ & $b$ & $V_{\mathrm{LSR}}$ & $d$  \\
 & & (h:m:s) & ($^\circ$:\arcmin:\arcsec) & ($^\circ$) & ($^\circ$) & (km s$^{-1}$) & (kpc) \\
(1) & (2) & (3) & (4) & (5) & (6) & (7) & (8)\\
\hline
C1 (MM9) & G028.37+00.07 (C) & 18:42:47.0 & -04:04:09 & 28.32450 & 0.06655 & 78.6 & 5.0\\
F1 (MM8) & G034.43+00.24 (F) & 18:53:16.6 & 01:26:10 & 34.41950 & 0.24583 & 57.1 & 3.7\\
F2 & G034.43+00.24 (F) & 18:53:19.1 & 01:26:54 & 34.43517 & 0.24217 & 57.1 & 3.7\\
G2 (MM2) & G034.77-00.55 (G) & 18:56:50.1 & 01:23:11 & 34.78117 & -0.56817 & 43.6 & 2.9\\
\hline
\end{tabular}
\tablefoot{Column 1 gives the name of the clump as denoted by \citet{Butler12} and, in parenthesis, the original name of the clump as assigned by \citet{Rathborne06}. The F2 clump was not given a designation by \citet{Rathborne06}. Column 2 gives the name of the IRDC in which the clump is embedded. Columns 3 and 4 give the right ascension and declination of the clump and Cols.\ 5 and 6 give the Galactic longitude and latitude of the clump, as given by \citet{Butler12}. Column 7 gives the velocity with respect to the local standard of rest of the parent IRDC, as given by \citet{Simon06}. Column 8 gives the kinematic distance of the cloud from \citet{Simon06}. While \citet{Kurayama11} find a 1.56 kpc distance for IRDC F, containing F1 and F2, \citet{Foster12} find a distance consistent with the kinematic distance for IRDC F, based upon extinction measurements. We thus elect to use the kinematic distances from \citet{Simon06} for all four clumps in this paper.}
\label{table:targets}
\end{center}
\end{minipage}
\end{table*}

\citet{Tan13} detect small, 0.1 pc sized cores in N$_2$D$^+$ emission within each of these clumps, with the C1 and G2 clumps harbouring two local maxima of N$_2$D$^+$ emission each. These individual N$_2$D$^+$ cores are denoted as C1-N, C1-S, G2-N, and G2-S. 

\begin{figure*}[htbp]
   \begin{minipage}[c][4 in][c]{\textwidth}
         \vspace*{\fill}
      \centering
 \includegraphics[height=4in]{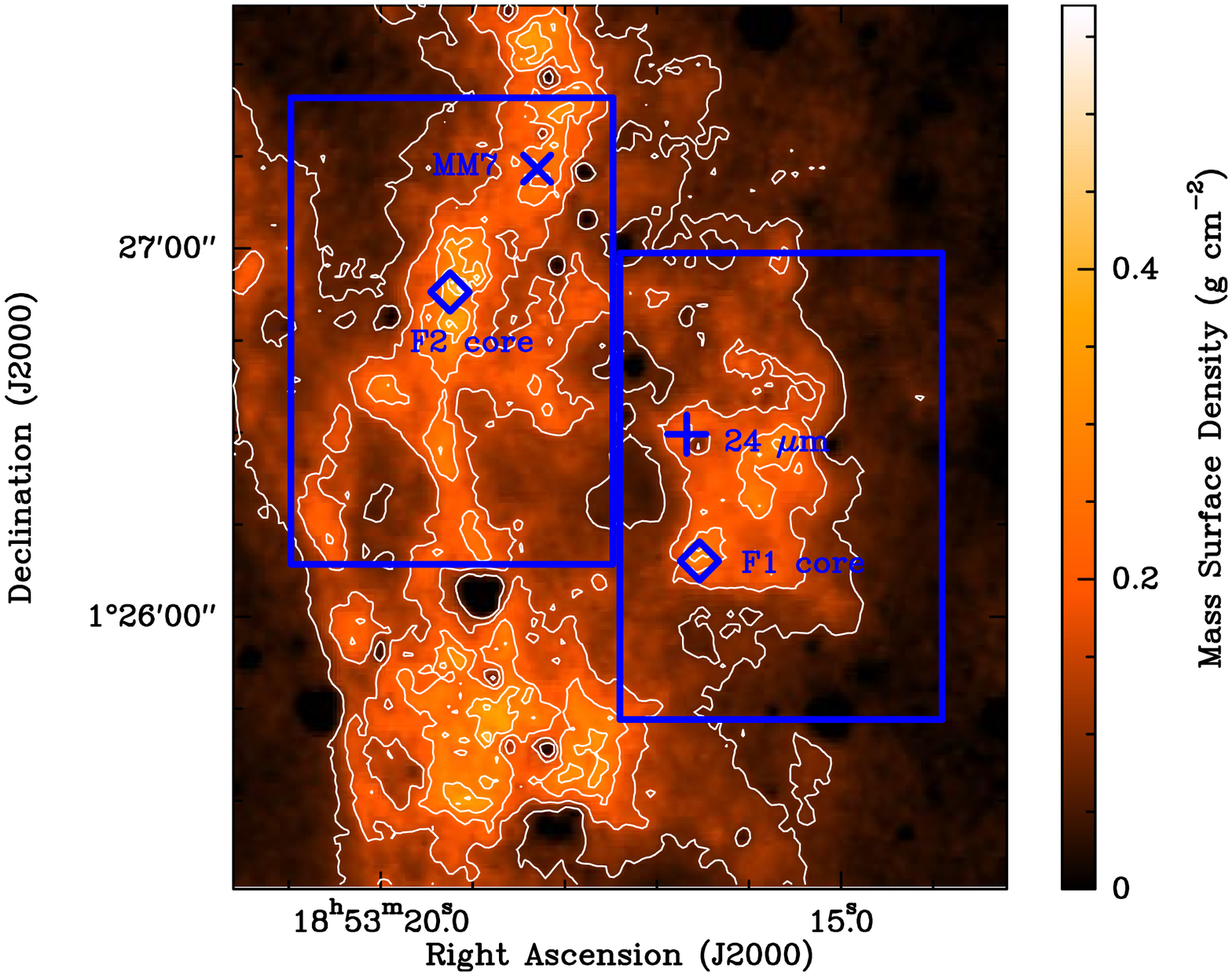}
\end{minipage}
   \begin{minipage}[c][3.5 in][c]{0.5 \textwidth}
      \vspace*{\fill}
      \centering
      \includegraphics[height=3in]{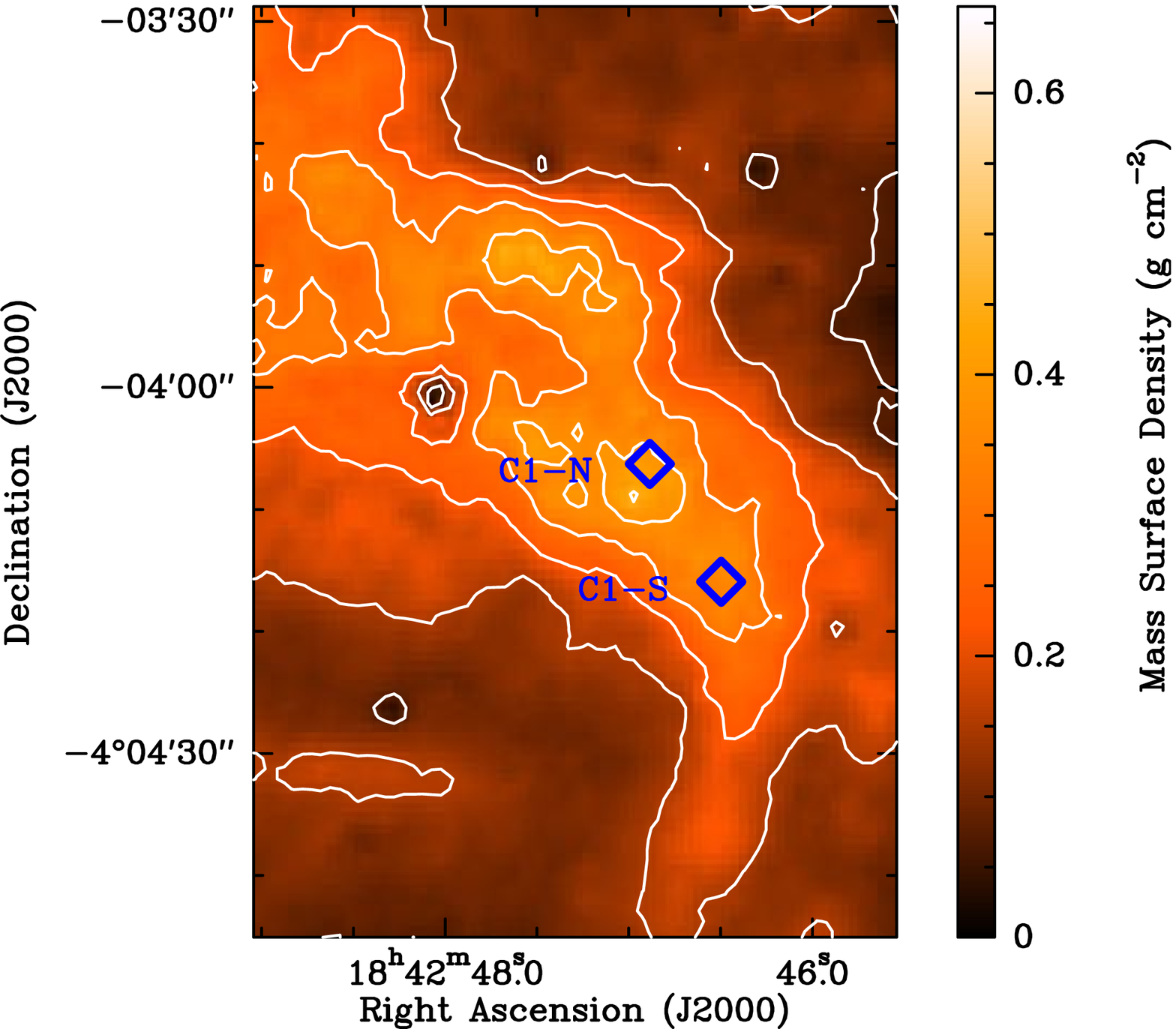}
   \end{minipage}
   \begin{minipage}[c][3.5 in][c]{0.5 \textwidth}
      \vspace*{\fill}
      \centering
       \includegraphics[height=3in]{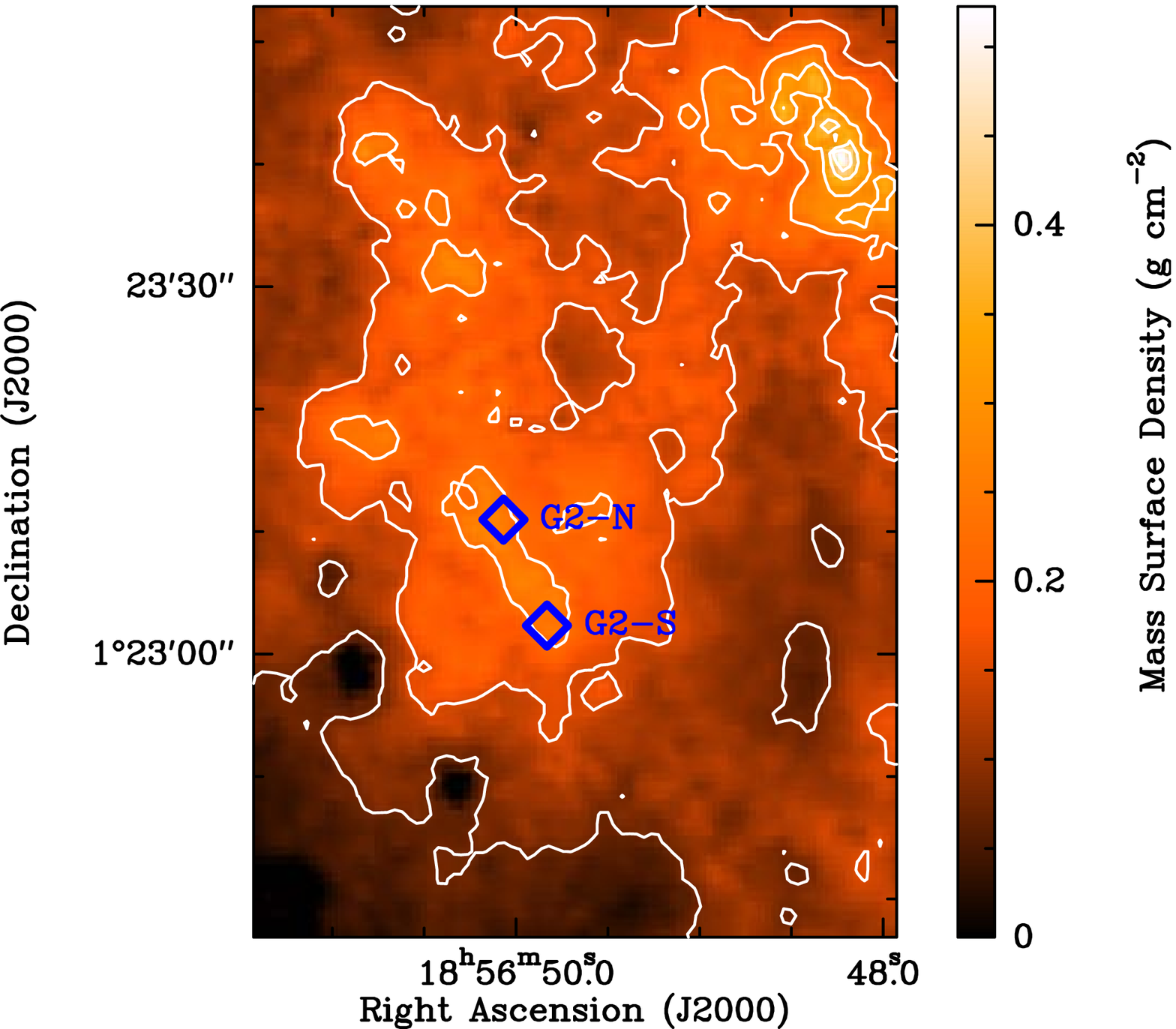}
   \end{minipage}
   \caption{Mass surface density derived by \citet{Butler12}. The {\it top, bottom left, and bottom right panels} show IRDCs F, C, and G, respectively. The contours start at 0.075 g cm$^{-2}$ ($A_{\mathrm{V}}$ of 17 mag) and increase by increments of 0.075 g cm$^{-2}$. The blue diamonds give the central locations of the cores as seen in N$_2$D$^+$ emission \citep{Tan13}. The regions shown in the {\it bottom two panels} (IRDCs C and G) are the regions mapped by {\it Herschel} in the CO $J$ = 9 $\rightarrow$ 8 line. The large blue rectangles in the {\it top panel} (IRDC F) show the areas mapped by {\it Herschel} in the CO $J$ = 9 $\rightarrow$ 8 line for the F1 and F2 clumps. In the {\it bottom left panel}, the mass surface density hole near 18$^h$:42$^m$:48.0$^s$, -4$^{\circ}$:04':00'' is due to the presence of an 8 micron source at this location preventing a reliable mass surface density from being determined from the extinction of the 8 micron Galactic background \citep{Butler14}. Such mass surface density holes, corresponding to locations of other 8 micron sources, can be seen in the other maps. In the {\it top panel}, the cross gives the location of the 24 micron source associated with the F1 clump \citep{Chambers09} and the ``X'' gives the location of the \citet{Rathborne06} MM7 clump.}
   \label{fig:sigmaradec}
\end{figure*}

Figure \ref{fig:sigmaradec} shows the mass surface densities of the four clumps, as derived from 8 micron extinction mapping \citep{Butler12}, and the central locations of the six cores. The contours start at 0.075 g cm$^{-2}$ (A$_V$ of 17 mag) and increase by increments of 0.075 g cm$^{-2}$, with the typical uncertainty being on the order of 30\% based upon uncertainties in the opacity per unit mass \citep{Butler09,Butler12}. The uncertainty based upon background fluctuations is of the order of 0.013 g cm$^{-2}$ \citep{Butler12}. This figure also shows the locations of the MM7 clump of \citet{Rathborne06} and the 24 micron source towards the edge of the F1 clump \citep{Chambers09}, which are discussed further in Sect.\ \ref{discussion}. 

The MM7 clump, at (18$^h$:53$^m$08.3$^s$, 1$^{\circ}$:27':13'') and first named by \citet{Rathborne06}, was previously called the F2 clump by \citet{Butler09}. When \citet{Butler12} updated the central locations of 42 of the \citet{Rathborne06} clumps by selecting the highest mass surface density peak within the clump areas defined by \citet{Rathborne06}, the F2 clump was the only clump for which \citet{Butler12} were unable to find a mass surface density peak within the clump boundary or within a few arcseconds of the boundary. The peak that \citet{Butler12} ended up assigning to F2 is $\sim10$ arcsec outside of the \citet{Rathborne06} boundary and 23 arcsec from the central location of MM7. Due to this position shift, we refer to the 0.5 pc scale mass surface density structure associated with the \citet{Butler12} F2 location as the F2 clump and refer to the structure associated with the \citet{Rathborne06} location as the MM7 clump. A closer examination of the \citet{Butler12} mass surface density map suggests that the F2 clump and MM7 clump are indeed different clumps (see Fig.\ \ref{fig:sigmaradec}).

\section{{\it HERSCHEL} OBSERVATIONS}
\label{Herschel}

\subsection{Reduction}
\label{reduction}

As part of the open time project OT2\_pcaselli\_8, roughly 1 arcminute square maps of the CO $J$ = 8 $\rightarrow$ 7, 9 $\rightarrow$ 8, and 10 $\rightarrow$ 9 line emission from the C1, F1, F2, and G2 clumps were obtained using the Heterodyne Instrument for the Far-Infrared (HIFI; \citealt{deGraauw10}) on board the {\it Herschel} Space Observatory \citep{Pilbratt10}. All of the observations were conducted in position switching mode with on-the-fly mapping. The off positions, in J2000 right ascension and declination, used for C1 and G2 were (18$^h$:42$^m$:28.18$^s$, -4$^\circ$:07\arcmin:22\arcsec) and (18$^h$:56$^m$:32.14$^s$, 1$^\circ$:24\arcmin:45\arcsec), respectively, while a common offset position of (18$^h$:53$^m$:26.18$^s$, 1$^\circ$:26\arcmin:16\arcsec) was used for F1 and F2. HIFI obtains data using both a Wide Band Spectrometer (WBS) and High Resolution Spectrometer (HRS). Since the WBS data have lower noise and the observed lines are quite weak, we will hereafter only use the WBS data. The noise in the WBS and HRS spectra are correlated, such that averaging the two spectra together would not produce any significant gain in the signal to noise of the lines. Further details of the observations, including the map centres, source times, beam sizes, observation bands, upper energy levels, rest frequencies, observation IDs, main beam efficiencies, and observation dates are given in Table \ref{table:observation setup}. 

\setlength{\tabcolsep}{5pt}
\begin{table*}
\begin{minipage}{\textwidth}
\caption{{\it Herschel} Observation Setup Parameters}
\begin{center}
\begin{tabular}{cccccccccccc}
\hline
\hline
Clump & Line & RA & Dec & $t$ & $HPBW$ & Band & $\nu$ & $E_{\mathrm{up}}$/$k$ & ID & $\eta_{\mathrm{mb}}$ & Date \\
 & & (h:m:s) & ($^\circ$:\arcmin:\arcsec) & (s) & (arcsec) & & (GHz) & (K) & & & (yy/mm/dd) \\
(1) & (2) & (3) & (4) & (5) & (6) & (7) & (8) & (9) & (10) & (11) & (12) \\
\hline
C1 & 8 $\rightarrow$ 7 & 18:42:47.29 & -4:04:06.1 & 265.5 & 23 & 3b & 921.79970 & 199 & 1342251119 & 0.628 & 12/09/18 \\
C1 & 9 $\rightarrow$ 8 & 18:42:47.29 & -4:04:06.1 & 1332 & 20 & 4a & 1036.91239 & 249 & 1342254432 & 0.635 & 12/10/24  \\
C1 & 10 $\rightarrow$ 9 & 18:42:47.29 & -4:04:06.1 & 1425.6 & 19 & 5a & 1151.98544& 304 & 1342253926 & 0.590 & 12/10/26 \\
F1 & 8 $\rightarrow$ 7 & 18:53:15.65 & 1:26:21.8  & 1110 & 23 & 3b & 921.79970 & 199 & 1342252136 & 0.628 & 12/10/03 \\
F1 & 9 $\rightarrow$ 8 & 18:53:15.65 & 1:26:21.8 & 1332 & 20 & 4a & 1036.91239 & 249 & 1342254434 & 0.635 & 12/10/25 \\
F1 & 10 $\rightarrow$ 9 & 18:53:15.65 & 1:26:21.8 & 1425.6 & 19 & 5a & 1151.98544& 304 & 1342253937& 0.590 & 12/10/26 \\
F2 & 8 $\rightarrow$ 7 & 18:53:19.22 & 1:26:47.2 & 814 & 23 & 3b & 921.79970 & 199 & 1342244947 & 0.628 & 12/04/25 \\
F2 & 9 $\rightarrow$ 8 & 18:53:19.22 & 1:26:47.2 & 1776 & 20 & 4a & 1036.91239 & 249 & 1342254433 & 0.635 &12/10/25 \\
F2 & 10 $\rightarrow$ 9 & 18:53:19.22 & 1:26:47.2 & 1425.6 & 19 & 5a & 1151.98544& 304 & 1342253936 & 0.590 & 12/10/26 \\
G2 & 8 $\rightarrow$ 7 & 18:56:49.67 & 1:23:15.4 & 265.5 & 23 & 3b & 921.79970 & 199 & 1342244946 & 0.628 &12/04/25 \\
G2 & 9 $\rightarrow$ 8 & 18:56:49.67 & 1:23:15.4 & 1332 & 20 & 4a & 1036.91239 & 249 & 1342254351 & 0.635 &12/11/03 \\
G2 & 10 $\rightarrow$ 9 & 18:56:49.67 & 1:23:15.4 & 1425.6 & 19 & 5a & 1151.98544& 304 & 1342253938 & 0.590 & 12/11/03 \\
\hline
\end{tabular}
\tablefoot{Column 1 gives the name of the clump observed while Col.\ 2 gives the $^{12}$CO transition observed. Columns 3 and 4 are the right ascension and declination, respectively, of the map centres. The on source integration time of the observation is given in Col.\ 5. The average half power beam width of the H and V polarizations is given in Col.\ 6 and the {\it Herschel} frequency band of the observation is given in Col.\ 7. Columns 8 and 9, respectively, give the rest frequency and upper level energy, in units of kelvin, of the observed transition from the Spectral Line Atlas of Interstellar Molecules (SLAIM; \citealt{Remijan07}). Columns 10 and 11 give the observation ID and the main beam conversion factor used. The main beam conversion factor used is the average of the appropriate values for the H and V polarizations based on the revised values of Jellema et al. (in prep.). Finally, Col.\ 12 gives the date the observation was conducted.}
\label{table:observation setup}
\end{center}
\end{minipage}
\end{table*}
\setlength{\tabcolsep}{6pt}

All of the {\it Herschel} data were reduced and processed using the {\it Herschel} Interactive Processing Environment (HIPE). During this reduction, multiple versions of HIPE between version 8 and 12.1.0 were used. The {\it Herschel} data were run through the default pipeline before being downloaded from the {\it Herschel} Science Archive. Standing waves were removed from the data and a second order polynomial was fit to the baseline. The H and V polarization data were combined and spectral cubes were created from the data. When making the spectral cubes, the doGridding task was initially allowed to choose the optimal map grid for the data. After examining the data to determine which maps show detections, the doGridding script was then rerun, on a clump by clump basis, on the original data to grid all of the data sets showing detections onto the grid map for the highest $J$ line with definitive detections. Due to the similarities in beam sizes for the different lines, the higher $J$ maps were not further smoothed to match the slightly larger beam size of the CO $J$ = 8 $\rightarrow$ 7 maps. The resulting data cubes were then converted from frequency space into velocity space, with respect to the local standard of rest, and then smoothed to a resolution of 2 km s$^{-1}$ (but with 1 km s$^{-1}$ sampling) from their original 1 MHz resolution ($\sim 0.3$ km s$^{-1}$). To convert between antenna temperatures and main beam temperatures, main beam efficiencies of 0.628, 0.635, and 0.590 were used for the CO $J$ = 8 $\rightarrow$ 7, 9 $\rightarrow$ 8, and 10 $\rightarrow$ 9 data, respectively (Jellema et al., in prep.). A forward efficiency of 0.96 is used for all observations \citep{Roelfsema12}. All intensities quoted in this paper are main beam temperatures, unless otherwise stated. 

To fit a Gaussian profile to each pixel of the data cubes, the HIPE multifit graphical user interface (GUI) was used. The pixel with the strongest apparent detection was first identified by eye and the GUI was used to fit a Gaussian to this spectrum. The best fit from this pixel was then used as the initial estimate for the automated fitting of the remainder of the pixels. Each fit was reviewed and any fits which appeared exceedingly poor were redone individually using the GUI. For any spectrum where it was ambiguous whether small features near the edges of the line were part of the line or just noise, the default fit from the multifit GUI was adopted. Since the hot gas component generated from turbulent dissipation should be relatively widespread across a molecular cloud, a map average spectrum was also created for each data set and a Gaussian was fit to these spectra using the HIPE fitting GUI. Since the CO $J$ = 10 $\rightarrow$ 9 line is not detected in any map, the spatially averaged CO $J$ = 10 $\rightarrow$ 9 spectra from all four regions were combined in IDL, after first shifting the velocity zero points of each spectra. For the C1, F1, and F2 data, the velocity zero point used was chosen to be the average central velocity of the spatially averaged CO $J$ = 8 $\rightarrow$ 7 and 9 $\rightarrow$ 8 spectra. For G2, the average of the N$_2$D$^+$ central velocities of the G2-N and G2-S cores was used \citep{Tan13}. 

While the GUI provides uncertainty estimates for most of the fit parameters, we choose to estimate the uncertainty in the integrated intensity from the equation
\begin{equation}
dI = RMS \times \sqrt{FWHM \times \delta v},
\label{eqn:di}
\end{equation} 
where dI is the uncertainty in the integrated intensity, RMS is the root mean square of the baseline, FWHM is the full width at half maximum of the line, and $\delta v$ is the velocity resolution. The RMS of the baseline is determined by applying the HIPE statistics task to the regions of the spectrum that appear to be devoid of any lines. 

The absolute flux calibration for these HIFI bands is of the order of 10\% \citep{Roelfsema12}, but has not been included in any below stated intensity uncertainties. The pointing accuracy of {\it Herschel} is of the order of 2'' \citep{Roelfsema12}. 

Any spectrum with an integrated intensity less than four times the uncertainty in the integrated intensity is classified as a non-detection. Spectra with integrated intensities between four and five times the uncertainty are classified as tentative detections and those spectra with integrated intensities greater than five times the uncertainty are classified as strong detections. Tentative and strong detections will jointly be referred to as detections and all further analysis in this paper will consider strong and tentative detections equally. We chose a cutoff at four times the uncertainty, rather than three times, because too many lines with integrated intensities between 3 and 4 times the uncertainty in the integrated intensity did not appear to be obvious, good quality fits by eye. 

For non-detections, an upper limit to the integrated intensity is assigned via
\begin{equation}
I_{\mathrm{lim}} = 4 \, RMS \times \sqrt{FWHM_{\mathrm{ave}} \times \delta v},
\label{eqn:rms}
\end{equation}
where $FWHM_{\mathrm{ave}}$ is the average full width at half maximum of all of the detected lines from the same map. Due to a lack of detected lines in the CO $J$ = 10 $\rightarrow$ 9 maps, the average FWHM of the CO $J$ = 9 $\rightarrow$ 8 map of the corresponding region is adopted in order to calculate the upper limits of the 10 $\rightarrow$ 9 spectra. For the G2 CO $J$ = 9 $\rightarrow$ 8 and 10 $\rightarrow$ 9 maps, where no lines are detected, the mean of the $FWHM_{\mathrm{ave}}$ values from the C1, F1, and F2 CO $J$ = 9 $\rightarrow$ 8 maps (5.4 km s$^{-1}$) is used to estimate the integrated intensity upper limit.  

\subsection{Results}
\label{herschel results}

\begin{figure*}
   \centering
   \begin{subfigure}[b]{0.33\textwidth}
   \centering
      \includegraphics[width=2.5 in]{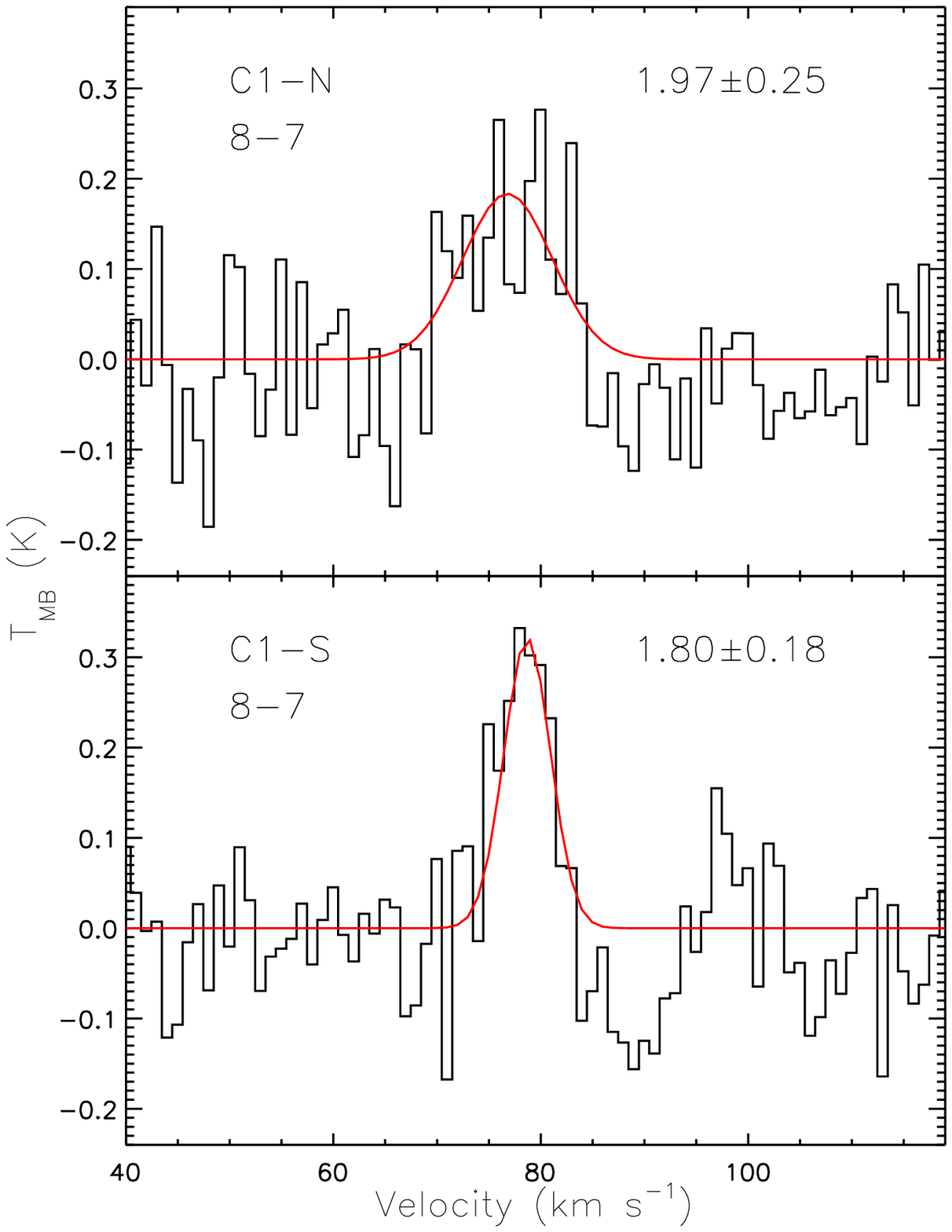}
   \end{subfigure}%
  \begin{subfigure}[b]{0.33\textwidth}
  \centering
      \includegraphics[width=2.5 in]{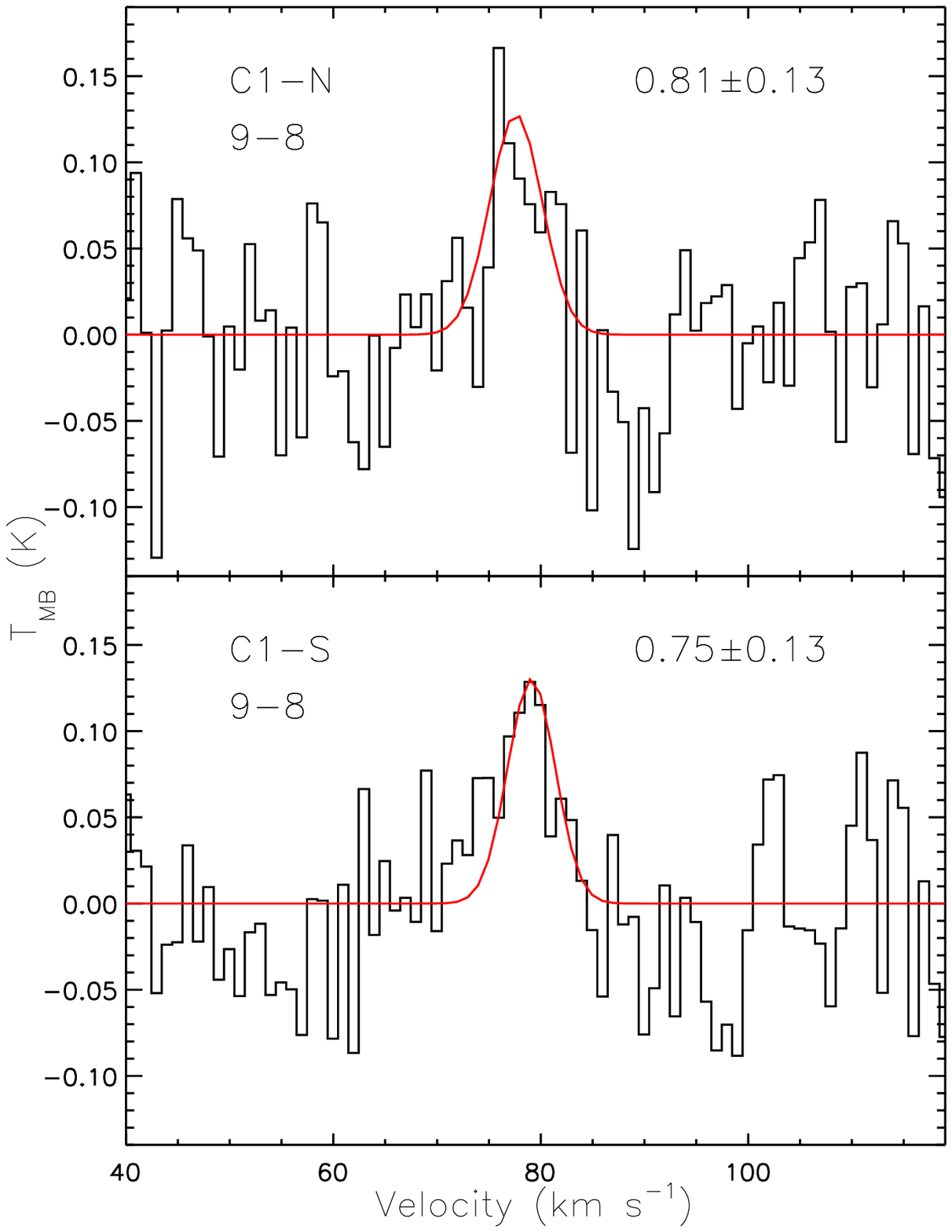}
   \end{subfigure}
   \begin{subfigure}[b]{0.33\textwidth}
   \centering
      \includegraphics[width=2.5 in]{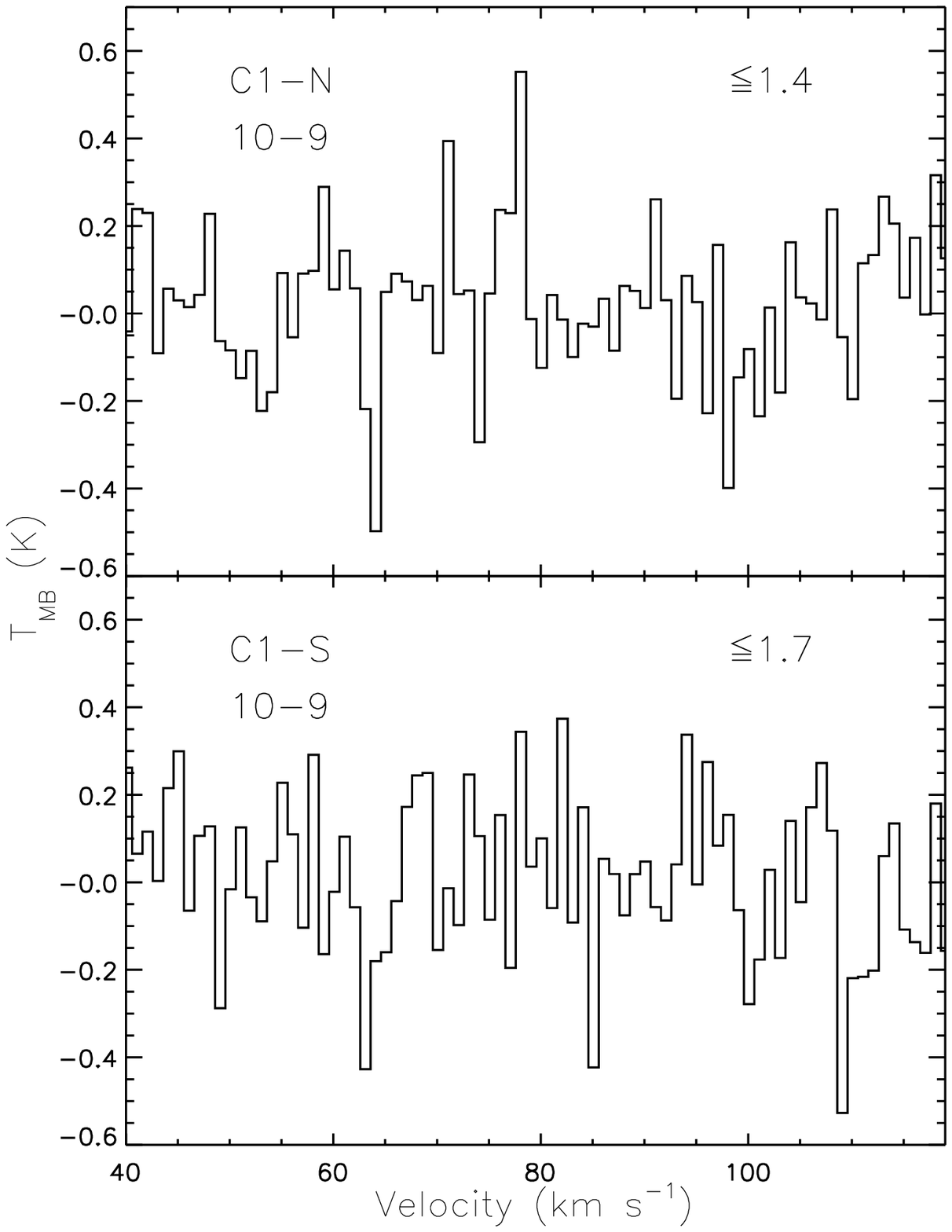}
   \end{subfigure}
   \caption{CO $J$ = 8 $\rightarrow$ 7, 9 $\rightarrow$ 8, and 10 $\rightarrow$ 9 lines ({\it left, middle, right}) towards the C1-N ({\it top}) and C1-S ({\it bottom}) cores. The best fitting Gaussians are overlaid in red. The integrated intensity of the fit and its uncertainty are given in K km s$^{-1}$ in the top right corner of each spectral box. For non-detections, the upper limit in K km s$^{-1}$ is instead given.}
   \label{fig:c1select}
\end{figure*}

\begin{figure*}
   \centering
   \begin{subfigure}[b]{0.33\textwidth}
   \centering
      \includegraphics[width=2.5 in]{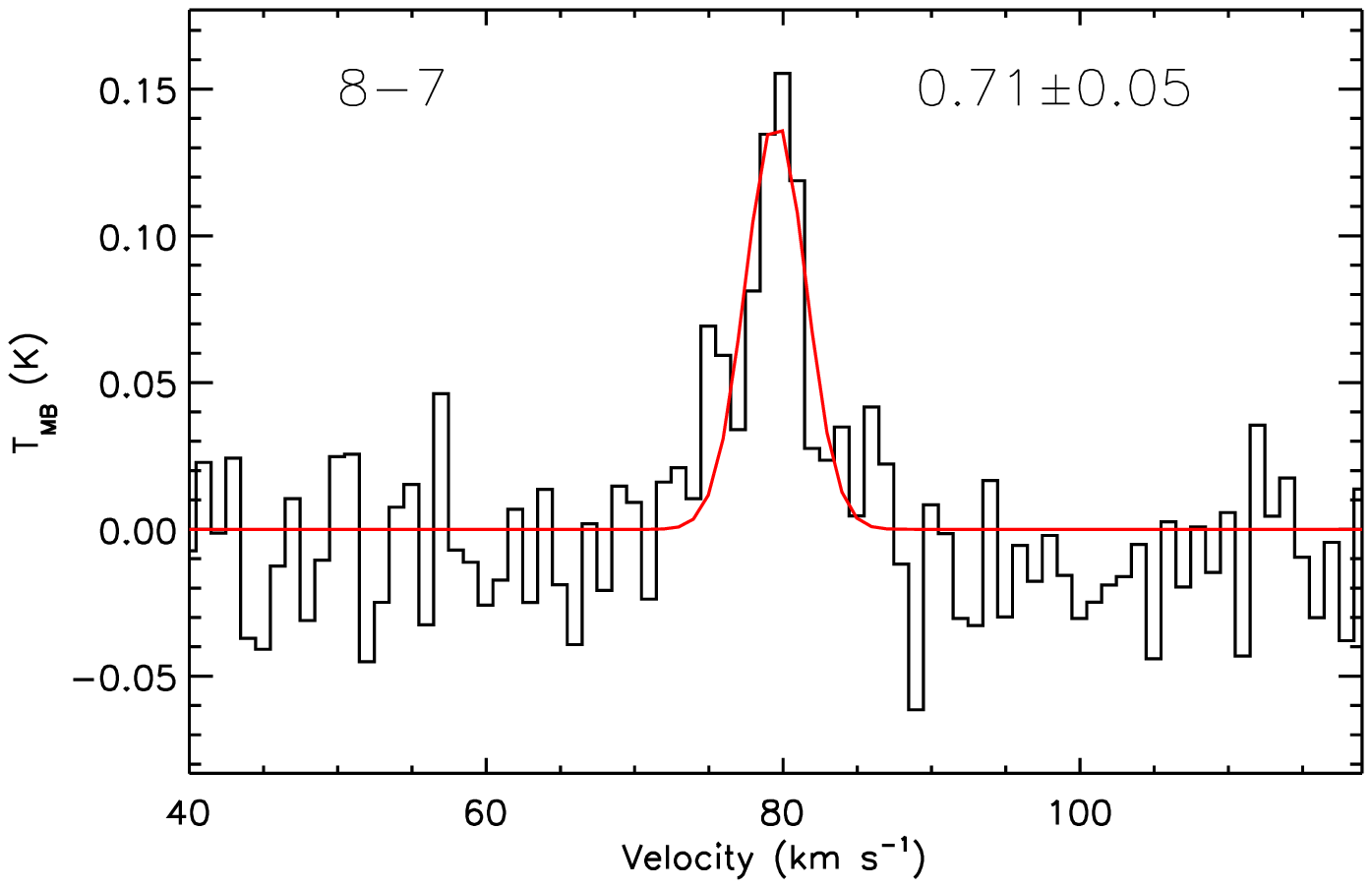}
   \end{subfigure}%
  \begin{subfigure}[b]{0.33\textwidth}
  \centering
      \includegraphics[width=2.5 in]{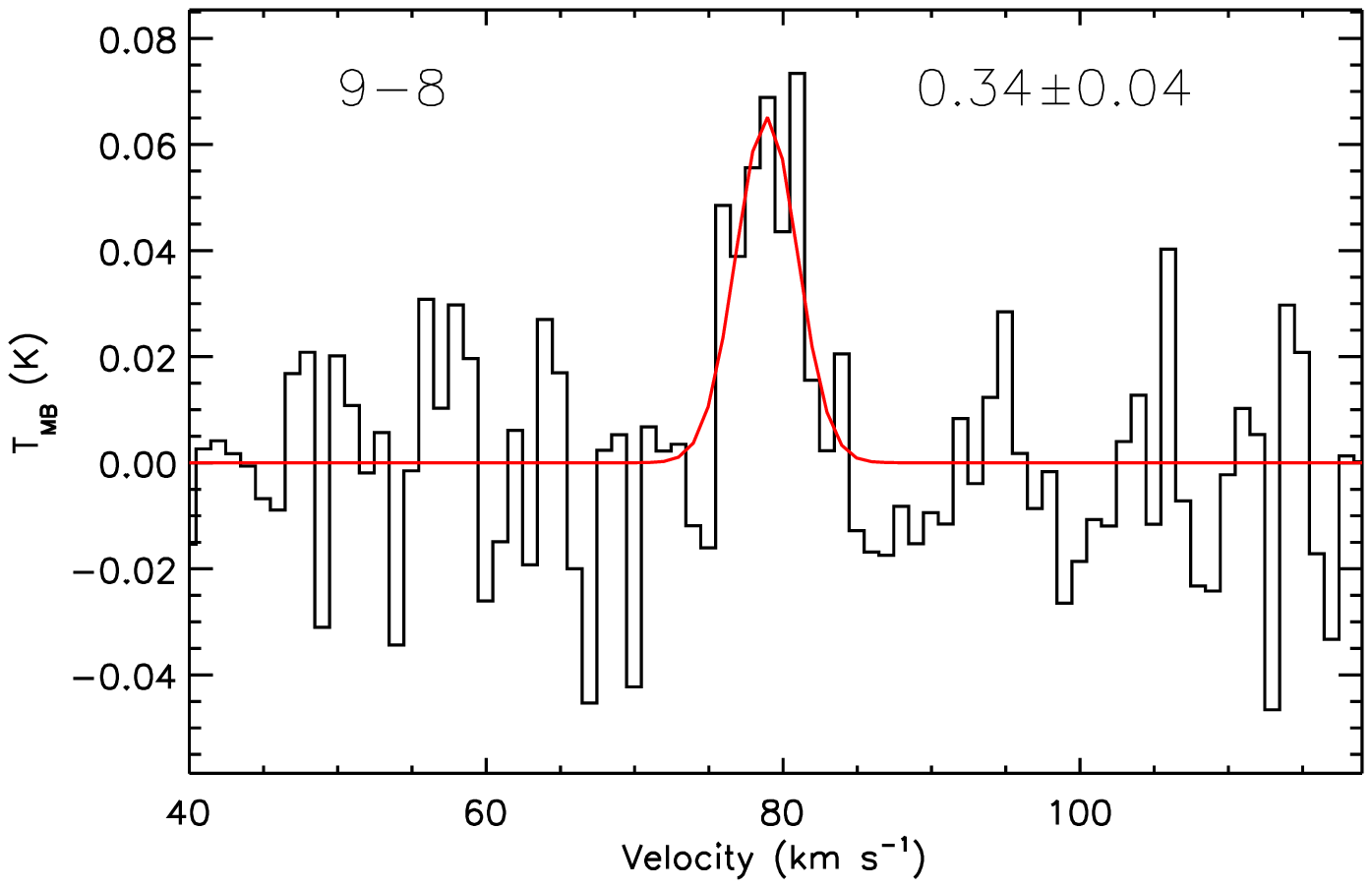}
   \end{subfigure}
   \begin{subfigure}[b]{0.33\textwidth}
   \centering
      \includegraphics[width=2.5 in]{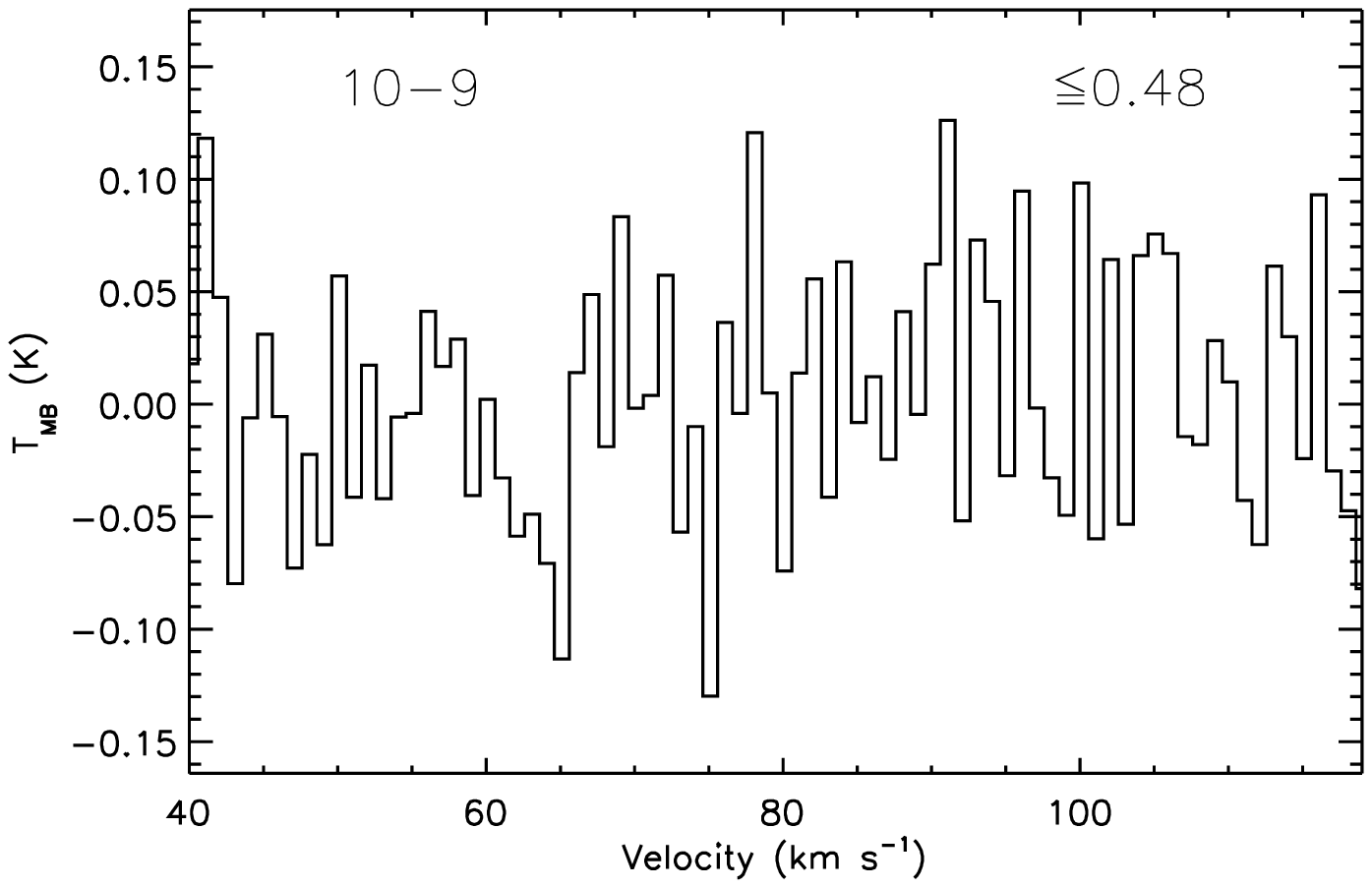}
   \end{subfigure}
   \caption{Spatially averaged spectra of the CO $J$ = 8 $\rightarrow$ 7, 9 $\rightarrow$ 8, and 10 $\rightarrow$ 9 lines ({\it left, middle, right}) towards the C1 clump. The best fitting Gaussians are overlaid in red. The integrated intensity of the fit and its uncertainty are given in K km s$^{-1}$ in the top right corner of each spectral box. For non-detections, the upper limit in K km s$^{-1}$ is instead given.}
   \label{fig:c1avefits}
\end{figure*}

Figure \ref{fig:c1select} shows the spectra and the associated Gaussian fits for the three observed lines towards the positions of the C1-N and C1-S cores. Figure \ref{fig:c1avefits} shows the spectra obtained by averaging all of the data, for each transition, over the C1 clump and the corresponding fits. These spectra are typical of strong detections (and non-detections in the case of the 10 $\rightarrow$ 9 lines). 

\begin{figure*}
   \centering
   \includegraphics[width=4.0in, angle=270]{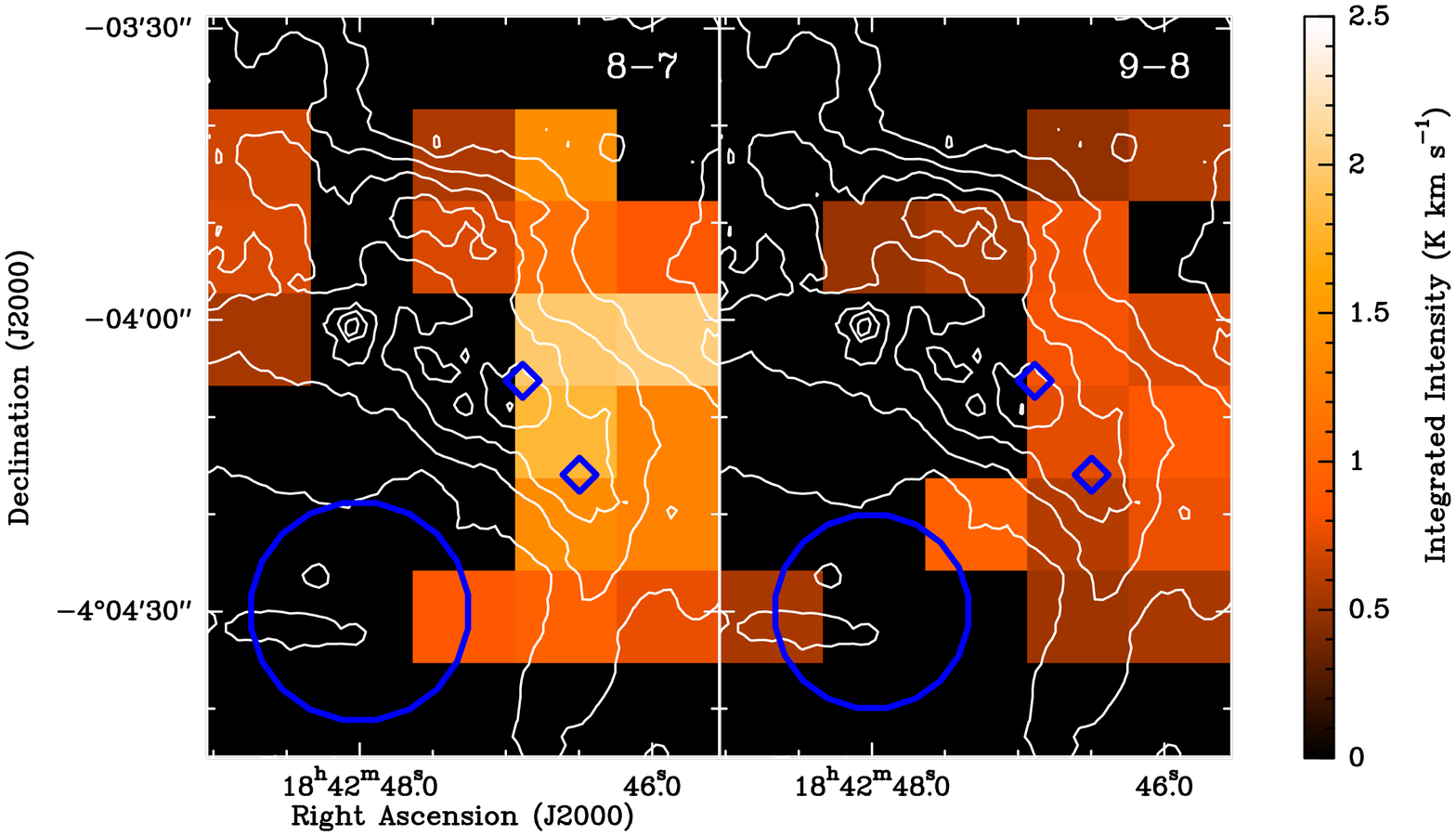}
   \caption{Integrated intensities of the CO $J$ = 8 $\rightarrow$ 7 ({\it left}) and CO $J$ = 9 $\rightarrow$ 8 ({\it right}) lines towards the C1 clump are shown in the colour scale. The contours are mass surface density from \citet{Butler12}, with the contours starting at 0.075 g cm$^{-2}$ and increasing by increments of 0.075 g cm$^{-2}$. The blue diamonds are the locations of the C1-N and C1-S cores, with the top left core being C1-N. The blue circles have diameters equal to the half power beam width of the {\it Herschel} beams.}
   \label{fig:cint}
\end{figure*}

\begin{figure*}
   \centering
   \includegraphics[width=4.0in, angle=270]{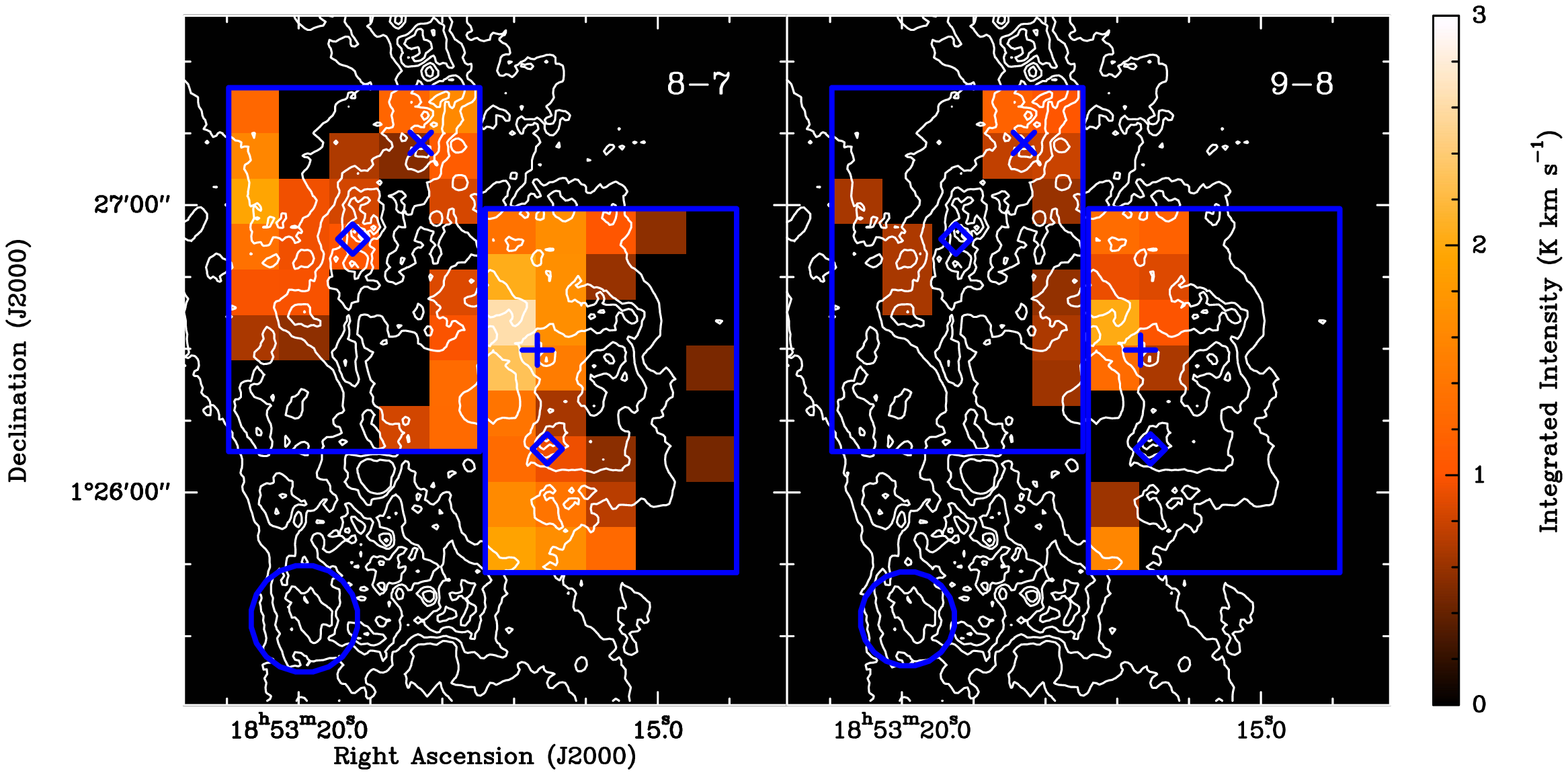}
   \caption{Integrated intensities of the CO $J$ = 8 $\rightarrow$ 7 ({\it left}) and CO $J$ = 9 $\rightarrow$ 8 ({\it right}) lines towards the F1 and F2 clumps are shown in the colour scale. The contours are mass surface density from \citet{Butler12}, with the contours starting at 0.075 g cm$^{-2}$ and increasing by increments of 0.075 g cm$^{-2}$. The blue diamonds are the locations of the F1 and F2 cores, with the top left core being F2. The large blue rectangles denote the area surveyed with {\it Herschel}. The cross gives the location of the 24 micron source associated with the F1 clump \citep{Chambers09} and the ``X'' gives the location of the \citet{Rathborne06} MM7 clump. The blue circles have diameters equal to the half power beam width of the {\it Herschel} beams.}
   \label{fig:fint}
\end{figure*}

\begin{figure*}
   \centering
   \begin{subfigure}[b]{0.4\textwidth}
   \centering
   \includegraphics[height=2.7in]{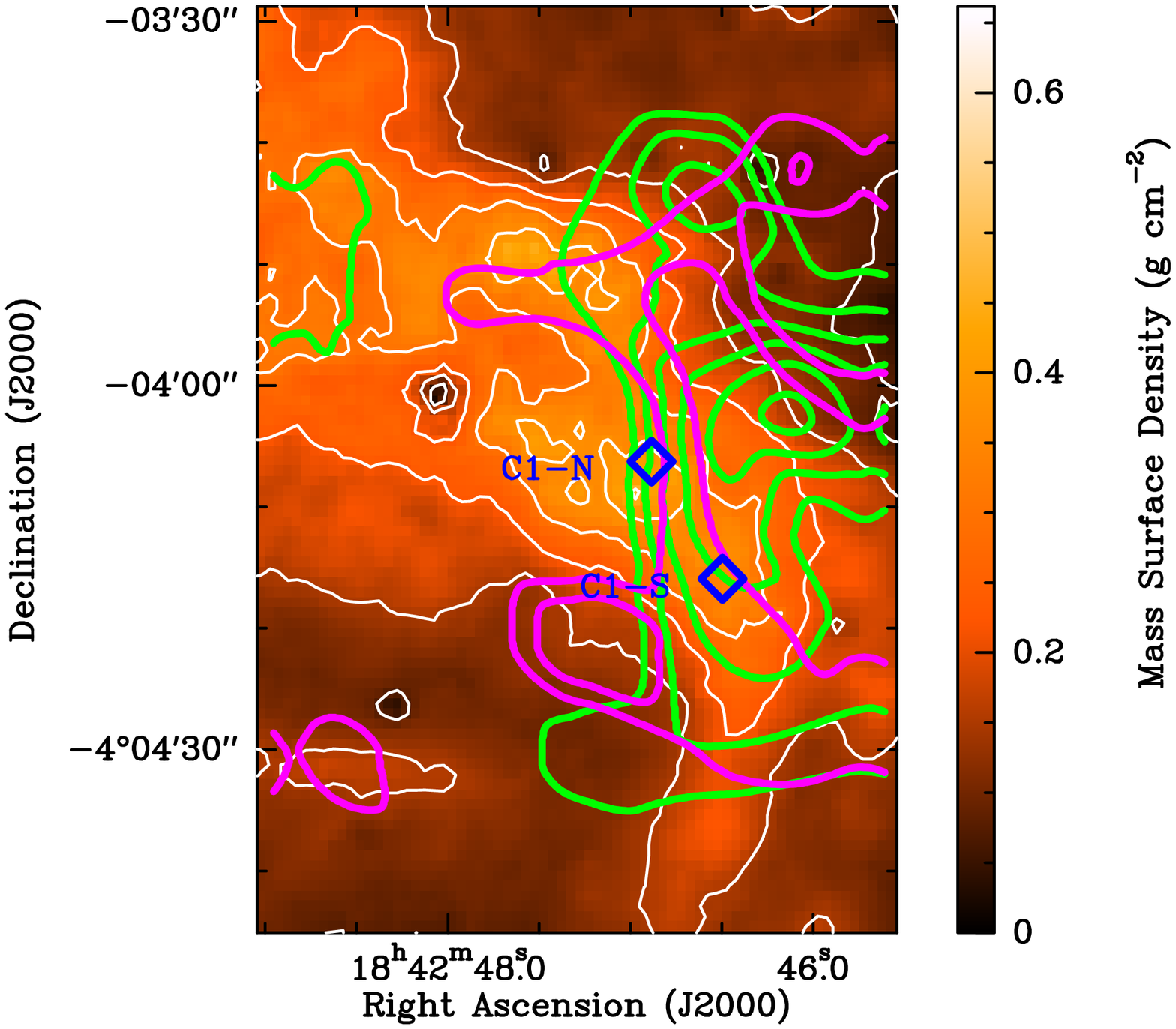}
   \end{subfigure}%
  \begin{subfigure}[b]{0.6\textwidth}
  \centering
   \includegraphics[height=2.7in]{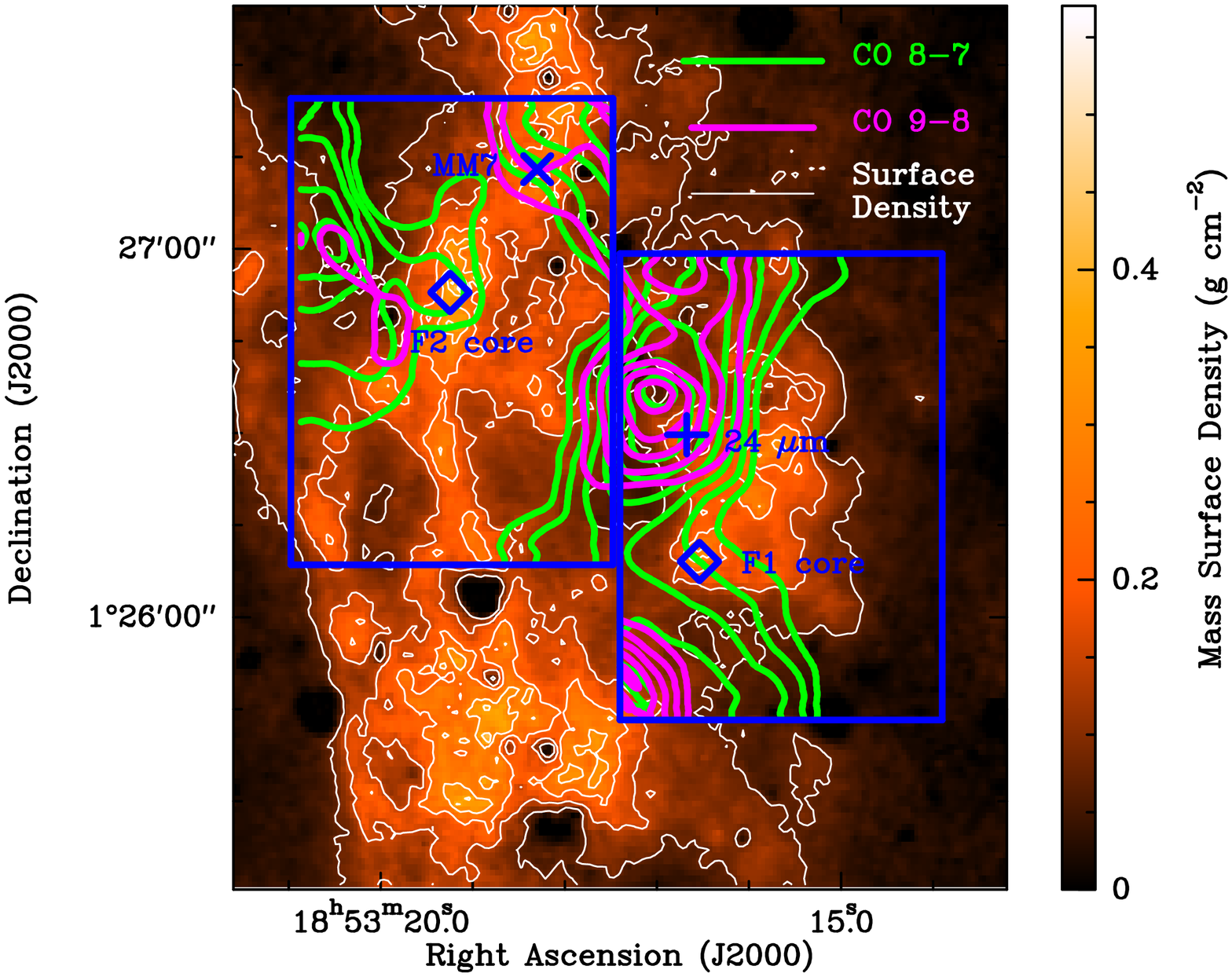}
   \end{subfigure}
   \caption{Integrated intensities of the CO $J$ = 8 $\rightarrow$ 7 (green) and CO $J$ = 9 $\rightarrow$ 8 (purple) lines are shown as the thick contours. The {\it left panel} shows the C1 clump while the {\it right panel} shows the F1 and F2 clumps. The contours start at four times the average uncertainty and increase in increments of two times the average uncertainty (see Col.\ 9 of Table \ref{table:Herschel fits}). For the {\it right panel}, the average uncertainty for each line was taken as the average value of $I_{lim}$ between the F1 and F2 maps. The colour scale and thin white contours are mass surface density from \citet{Butler12}, with the contours starting at 0.075 g cm$^{-2}$ and increasing by increments of 0.075 g cm$^{-2}$. The blue diamonds are the locations of the C1-N, C1-S, F1, and F2 cores. The cross gives the location of the 24 micron source associated with the F1 clump \citep{Chambers09} and the ``X'' gives the location of the \citet{Rathborne06} MM7 clump. The large blue rectangles on the {\it right panel} denote the area surveyed with {\it Herschel}, while the {\it left panel} shows the entire {\it Herschel} survey region around the C1 clump.}
   \label{fig:intcontour}
\end{figure*}

Figures \ref{fig:cint} and \ref{fig:fint} show the integrated intensities of the CO $J$ = 8 $\rightarrow$ 7 and 9 $\rightarrow$ 8 transitions towards the F1, F2, and C1 clumps. Given the close proximity of F1 and F2, both clumps are placed on the same figure. The G2 clump is not shown, as the CO $J$ = 8 $\rightarrow$ 7 line is only tentatively detected in one location within G2 and the CO $J$ = 9 $\rightarrow$ 8 line is not detected towards any location within G2. Maps for the CO $J$ = 10 $\rightarrow$ 9 transition are also not shown as this line is not detected towards any location. Figure \ref{fig:intcontour} shows the integrated intensities of the CO $J$ = 8 $\rightarrow$ 7 and 9 $\rightarrow$ 8 transitions towards the F1, F2, and C1 clumps plotted as contours overtop of the mass surface densities of these regions, for easier comparison between the different data sets. 

The results of the Gaussian fits to the individual pixels are given in Table \ref{table:Herschel fits}. This table lists the fraction of pixels with tentative and strong detections; the maximum peak intensity, integrated intensity, and FWHM; the average peak intensity, integrated intensity, FWHM, and central velocity with respect to the local standard of rest; and the minimum FWHM. Table \ref{table:wbs average fits} gives the best fit parameters for the spatially averaged spectra. Note that in Table \ref{table:Herschel fits}, the typical upper limit to the integrated intensities is given ($I_{lim}$, see Eq.\ (\ref{eqn:rms})), which is equal to four times the typical integrated intensity uncertainty, while in Table \ref{table:wbs average fits}, it is the integrated intensity uncertainty that is instead quoted ($dI$, see Eq.\ (\ref{eqn:di})).

\setlength{\tabcolsep}{2pt}
\begin{table*}
\tiny
\begin{minipage}{\textwidth}
\caption{Fitting Results for the {\it Herschel} Data}
\begin{center}
\begin{tabular}{ccccccccccccc}
\hline
\hline
Clump & Line & $f_{\mathrm{tent}}$ & $f_{\mathrm{strong}}$ & $T_{\mathrm{MB,max}}$ & $T_{\mathrm{MB,ave}}$ & $I_{\mathrm{max}}$ & $I_{\mathrm{ave}}$ & $I_{\mathrm{lim}}$ & $FWHM_{\mathrm{max}}$ & $FWHM_{\mathrm{min}}$ & $FWHM_{\mathrm{ave}}$ & $V_{\mathrm{LSR}}$ \\
 & & & & (K) & (K) & (K km s$^{-1}$) & (K km s$^{-1}$) & (K km s$^{-1}$) & (km s$^{-1}$) & (km s$^{-1}$) & (km s$^{-1}$) & (km s$^{-1}$) \\
(1) & (2) & (3) & (4) & (5) & (6) & (7) & (8) & (9) & (10) & (11) & (12) & (13) \\
\hline
C1 & 8 $\rightarrow$ 7 & 4 (10.0\%) & 13 (32.5\%) & 0.58 & 0.27 & 2.03 & 1.11 & 0.67 & 10.2 & 2.1 & 4.6 & 79.2 \\
C1 & 9 $\rightarrow$ 8 & 7 (17.5\%) & 8 (20.0\%) & 0.21 & 0.14 & 0.96 & 0.67 & 0.50 & 5.9 & 3.5 & 4.6 & 79.4 \\
C1 & 10 $\rightarrow$ 9 & 0 & 0 & ... & ... & ... & ... & $\le$1.75 & ... & ... & ... & ... \\
F1 & 8 $\rightarrow$ 7 & 5 (12.5\%) & 19 (47.5\%) & 0.41 & 0.23 & 2.63 & 1.31 & 0.62 & 8.7 & 2.3 & 5.4 & 56.3 \\
F1 & 9 $\rightarrow$ 8 & 3 (7.5\%) & 7 (17.5\%) & 0.21 & 0.15 & 2.03 & 1.14 & 0.75 & 9.5 & 5.1 & 6.9 & 55.5 \\
F1 & 10 $\rightarrow$ 9 & 0 & 0 & ... & ... & ... & ... & $\le$2.16 & ... & ... & ... & ... \\
F2 & 8 $\rightarrow$ 7 & 5 (12.5\%) & 18 (45.0\%) & 0.33 & 0.19 & 1.96 & 1.05 & 0.67 & 7.4 & 3.0 & 5.4 & 58.0 \\
F2 & 9 $\rightarrow$ 8 & 3 (7.5\%) & 8 (20.0\%) & 0.26 & 0.15 & 1.16 & 0.75 & 0.54 & 5.8 & 3.8 & 4.8 & 57.7 \\
F2 & 10 $\rightarrow$ 9 & 0 & 0 & ... & ... & ... & ... & $\le$2.12 & ... & ... & ... & ... \\
G2 & 8 $\rightarrow$ 7 & 1 (2.9\%) & 0 (0\%) & 0.18 & 0.18 & 0.72 & 0.72 & 0.64 & 3.8 & 3.8 & 3.8 & 62.0 \\
G2 & 9 $\rightarrow$ 8 & 0 & 0 & ... & ... & ... & ... & $\le$0.60 & ... & ... & ... & ... \\
G2 & 10 $\rightarrow$ 9 & 0 & 0 & ... & ... & ... & ... & $\le$1.95 & ... & ... & ... & ... \\
\hline
\end{tabular}
\tablefoot{Column 1 gives the name of the clump observed while Col.\ 2 gives the CO transition observed. Columns 3 and 4 give the number of pixels in which there are tentative and strong detections, respectively. The number in parentheses is the corresponding percentage of pixels in the entire map. The maximum peak intensity and mean peak intensity of all of the detections are given in Cols.\ 5 and 6, while the maximum and mean integrated intensities of all of the detections are given in Cols.\ 7 and 8, respectively. Column 9 gives four times the average uncertainty in the integrated intensities of all of the detections, which gives a measure of the typical upper limits for any non-detections. For maps with no detections, Col.\ 9 gives four times the average uncertainty of all of the pixels. Columns 10-12 give the maximum, minimum, and mean FWHM of all detections, respectively. Column 13 gives the mean central velocity with respect to the local standard of rest of all of the detections.}
\label{table:Herschel fits}
\end{center}
\end{minipage}
\end{table*}
\setlength{\tabcolsep}{6pt}

\begin{table*}
\begin{minipage}{\textwidth}
\caption{Spatially Averaged Fit Parameters}
\begin{center}
\begin{tabular}{cccccc}
\hline
\hline
Clump & Line & $T_{\mathrm{MB}}$ & I & FWHM & $V_{\mathrm{LSR}}$ \\
 & & (K) & (K km s$^{-1}$) & (km s$^{-1}$) & (km s$^{-1}$) \\
(1) & (2) & (3) & (4) & (5) & (6) \\
\hline
C1 & 8 $\rightarrow$ 7 & 0.14 (0.02) & 0.71 (0.05) & 4.8 (0.6) & 79.5 (0.2) \\
C1 & 9 $\rightarrow$ 8 & 0.065 (0.012) & 0.34 (0.04) & 4.9 (1.0) & 78.9 (0.4) \\
C1 & 10 $\rightarrow$ 9 & ... & ... (0.12) & ... & ... \\
F1 & 8 $\rightarrow$ 7 & 0.13 (0.02) & 0.92 (0.08) & 6.7 (0.9) & 55.6 (0.4) \\
F1 & 9 $\rightarrow$ 8 & 0.054 (0.009) & 0.35 (0.04) & 6.1 (1.4) & 55.9 (0.6) \\
F1 & 10 $\rightarrow$ 9 & ... & ... (0.15) & ... & ... \\
F2 & 8 $\rightarrow$ 7 & 0.14 (0.02) & 0.69 (0.08) & 4.7 (0.9) & 57.9 (0.4) \\
F2 & 9 $\rightarrow$ 8 & 0.056 (0.010) & 0.36 (0.04) & 6.0 (1.3) & 58.4 (0.5) \\
F2 & 10 $\rightarrow$ 9 & ... & ... (0.12) & ... & ... \\
G2 & 8 $\rightarrow$ 7 & 0.05 (0.01) & 0.31 (0.06) & 5.7 (1.6) & 59.7 (0.7) \\
G2 & 9 $\rightarrow$ 8 & ... & ... (0.04) & ... & ... \\
G2 & 10 $\rightarrow$ 9 & ... & ... (0.14) & ... & ... \\
\hline
\end{tabular}
\tablefoot{Column 1 gives the name of the clump observed while Col.\ 2 gives the CO transition observed. Columns 3-6 give the best fit parameters for the spatially averaged spectrum from each map. The parentheses give the uncertainties for each parameter, as given by the Gaussian fitting routine, except for the integrated intensity, where the uncertainty is given by Eq.\ (\ref{eqn:di}). Column 3 gives the peak intensity while column 4 gives the integrated intensity. The FWHM is given in Col.\ 5. The central velocity with respect to the local standard of rest is given in Col.\ 6.}
\label{table:wbs average fits}
\end{center}
\end{minipage}
\end{table*}

In the C1, F1, and F2 maps, the CO $J$ = 8 $\rightarrow$ 7 line is detected in roughly half of the pixels and the 9 $\rightarrow$ 8 line is detected in over 25\% of the pixels. In the G2 map, the CO $J$ = 8 $\rightarrow$ 7 line is only detected in one pixel and in the spatial average. Furthermore, in the G2 map, the CO $J$ = 9 $\rightarrow$ 8 line is not detected at any location and is not detected in the spatial average.  The CO $J$ = 10 $\rightarrow$ 9 is not detected in any of the four clumps, nor in any of the spatially averaged spectra of each clump. 

No line is detected in the CO $J$ = 10 $\rightarrow$ 9 spectra obtained from combining the spatial averages of all four mapped regions, with an upper limit on this CO $J$ = 10 $\rightarrow$ 9 integrated intensity of 0.26 K km s$^{-1}$. It should be noted, however, that given the differences in the CO $J$ = 9 $\rightarrow$ 8 and 8 $\rightarrow$ 7 integrated intensities between the four regions, in particular between the G2 clump and the other three clumps, it is by no means obvious that the CO $J$ = 10 $\rightarrow$ 9 intensity should be similar between the four regions. Thus, the CO $J$ = 10 $\rightarrow$ 9 integrated intensity may still be slightly larger than 0.26 K km s$^{-1}$ in any particular clump. 

The C1, F1, and F2 maps have reasonably similar integrated intensity ranges for detections, with the detections of the CO $J$ = 8 $\rightarrow$ 7 and 9 $\rightarrow$ 8 lines having integrated intensities ranging from 0.45 to 2.63 K km s$^{-1}$ and from 0.5 to 2 K km s$^{-1}$, respectively. The average integrated intensity of a detection of the CO $J$ = 8 $\rightarrow$ 7 line varies from 1.11 to 1.31 to 1.05 K km s$^{-1}$ between the C1, F1, and F2 maps, and the average detection of the 9 $\rightarrow$ 8 line for these three regions changes from 0.67 to 1.14 to 0.75 K km s$^{-1}$. The integrated intensity of the spatial averages tends to be approximately a factor of two lower than the average detection from a single pixel and the emission is clearly spatially inhomogeneous within each clump. The pixel with both the largest CO $J$ = 8 $\rightarrow$ 7 and 9 $\rightarrow$ 8 integrated intensity lies just to the northeast of the 24 micron source in the F1 clump. 

\begin{figure*}
   \centering
   \begin{subfigure}[b]{0.5\textwidth}
   \centering
      \includegraphics[height=2.7in]{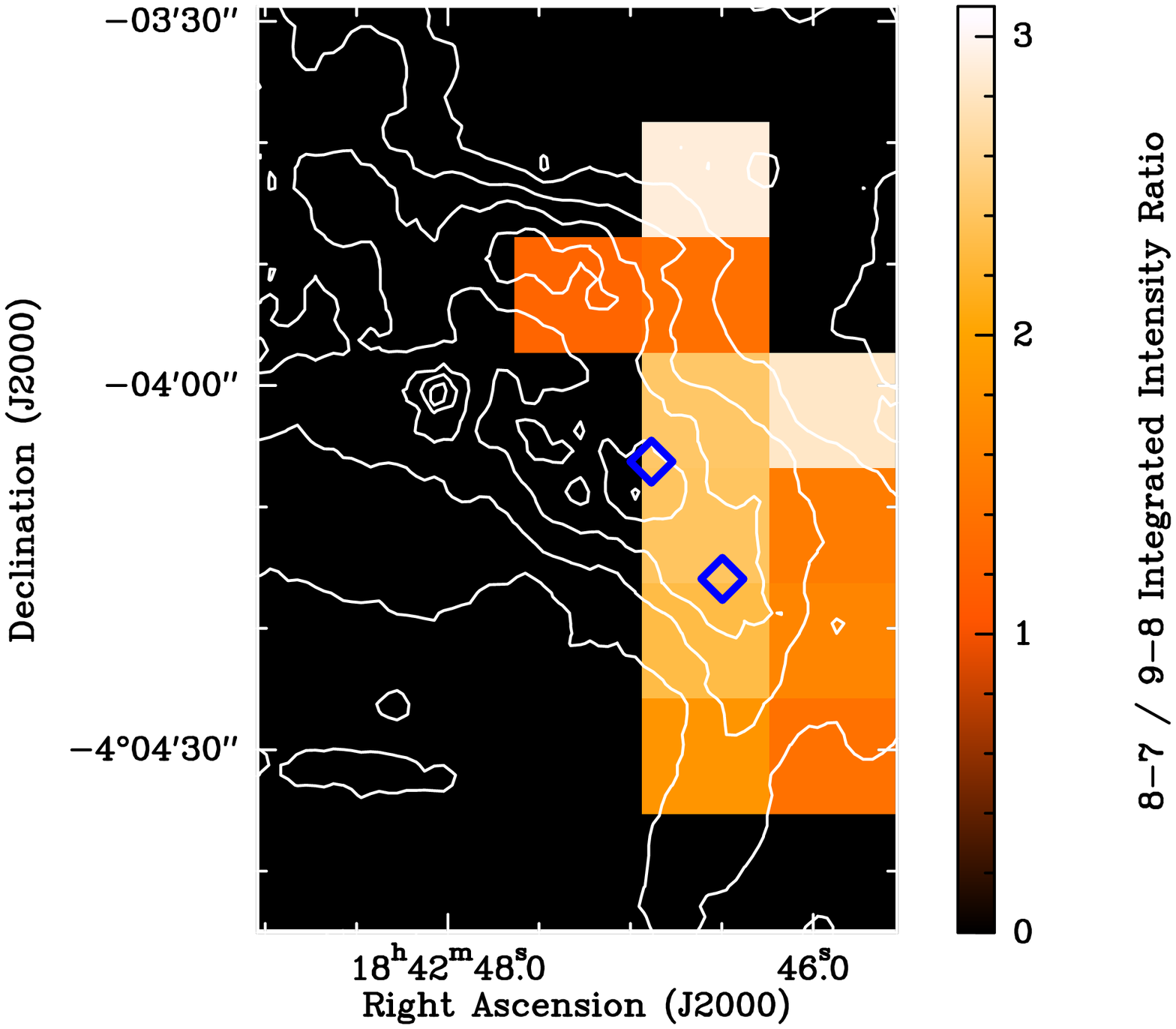}
   \end{subfigure}%
  \begin{subfigure}[b]{0.5\textwidth}
  \centering
      \includegraphics[height=2.7in]{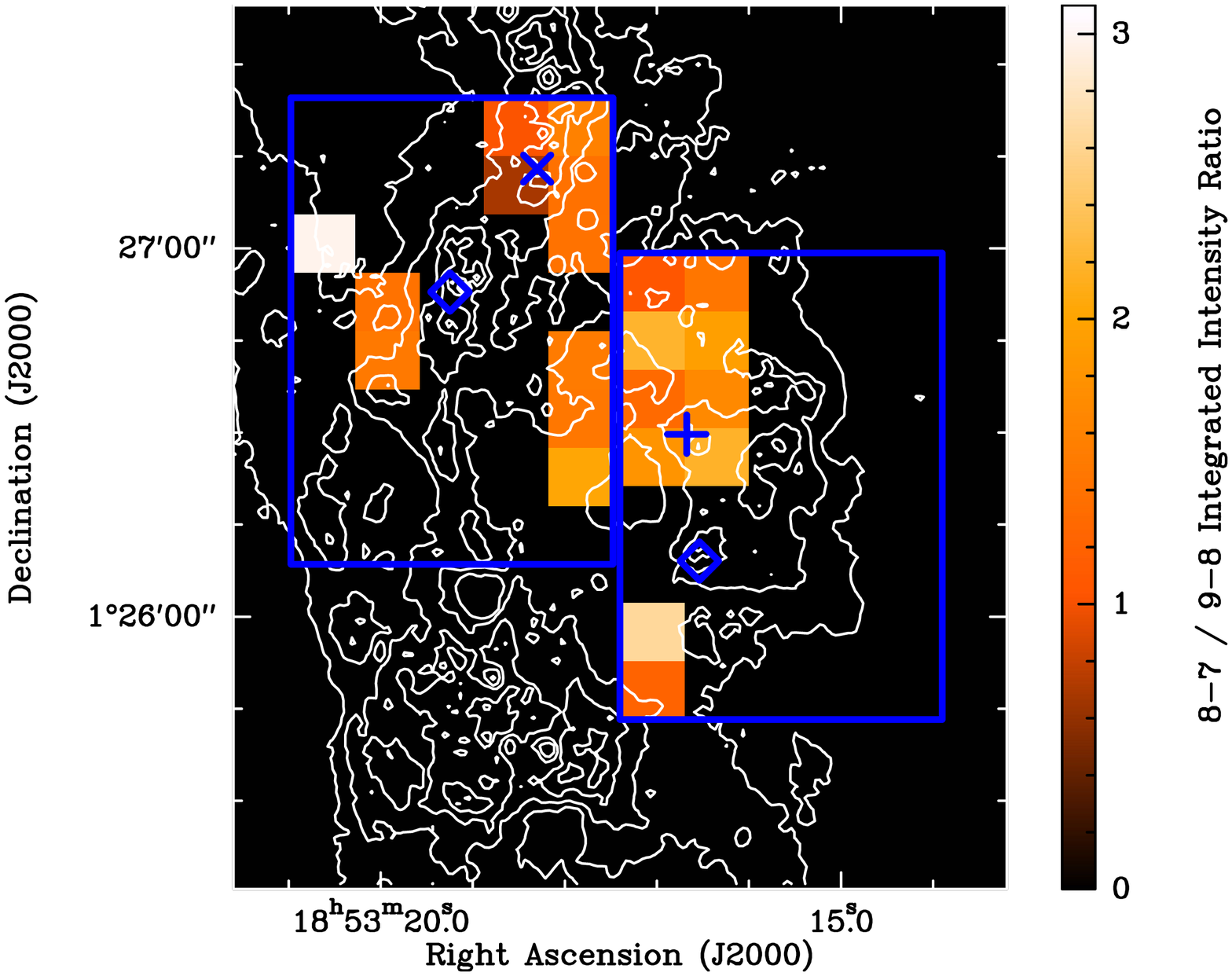}
   \end{subfigure}
   \caption{Ratios of the integrated intensities of the CO $J$ = 8 $\rightarrow$ 7 to 9 $\rightarrow$ 8 lines for locations where both lines are detected are shown in the colour scale. The {\it left panel} shows the C1 clump while the {\it right panel} shows the F1 and F2 clumps. The contours are mass surface density from \citet{Butler12}, with the contours starting at 0.075 g cm$^{-2}$ and increasing by increments of 0.075 g cm$^{-2}$. The blue diamonds are the locations of the C1-N, C1-S, F1, and F2 cores. See Fig.\ \ref{fig:sigmaradec} for the labels of these cores. The cross gives the location of the 24 micron source associated with the F1 clump \citep{Chambers09} and the ``X'' gives the location of the \citet{Rathborne06} MM7 clump. The large blue rectangles denote the area surveyed with {\it Herschel}.}
   \label{fig:ratios}
\end{figure*}

Figure \ref{fig:ratios} shows the ratio of the CO $J$ = 8 $\rightarrow$ 7 to 9 $\rightarrow$ 8 integrated intensities for all locations with detections in both lines. Table \ref{table:ratios} gives the minimum, maximum, and average integrated intensity ratio for the C1, F1, and F2 maps. The CO $J$ = 8 $\rightarrow$ 7 to 9 $\rightarrow$ 8 integrated intensity ratio ranges from 0.7 to 3.0, with a weighted average ratio of 2.0, 1.7, and 1.6 for all the points in clumps C1, F1, and F2, respectively, where both lines are detected. This weighted average is found by weighting the ratios by the inverse of their fractional uncertainties, which are typically on the order of 20-25\%. The ratios of the integrated intensities from the spatially averaged WBS spectra are 2.1 $\pm$ 0.3, 2.6 $\pm$ 0.4, and 1.9 $\pm$ 0.3 for clumps C1, F1, and F2. 

The ratio of the peak intensity of the CO $J$ = 8 $\rightarrow$ 7 and 9 $\rightarrow$ 8 lines are reasonably similar, with weighted averages of 2.1, 2.0, and 1.4 for all of the points with detections in both transitions and ratios of 2.1 $\pm$ 0.5, 2.4 $\pm$ 0.5, and 2.4 $\pm$ 0.6 for the spatially averaged WBS spectra (for C1, F1, and F2 respectively). 

\begin{table}
\caption{CO $J$ = 8 $\rightarrow$ 7 to 9 $\rightarrow$ 8 Integrated Intensity Ratios}
\begin{center}
\begin{tabular}{cccccc}
\hline
\hline
Clump & $f_{\mathrm{ratio}}$ & min & max & ave & WBS \\
(1) & (2) & (3) & (4) & (5) & (6) \\
\hline
C1 & 11 (27.5\%) & 1.2 & 2.9 & 2.0 & 2.1 (0.3)\\
F1 & 10 (25.0\%) & 1.1 & 2.6 & 1.7 & 2.6 (0.4)\\
F2 & 11 (27.5\%) & 0.7 & 3.0 & 1.6 & 1.9 (0.3)\\
\hline
\end{tabular}
\tablefoot{Column 1 gives the name of the clump observed. Column 2 gives the number of pixels where both the CO $J$ = 8 $\rightarrow$ 7 and 9 $\rightarrow$ 8 lines are detected while the value in parentheses gives the corresponding fraction of pixels with detections in both lines. Columns 3-5 give, respectively, the minimum, maximum, and weighted average ratio of the CO $J$ = 8 $\rightarrow$ 7 to 9 $\rightarrow$ 8 integrated intensities. The integrated intensity ratio of these two lines from the spatially averaged spectra is given in Col.\ 6, with the number in parentheses being the uncertainty in the ratio.}
\label{table:ratios}
\end{center}
\end{table}

In the F1, F2, and G2 maps, detection of the CO $J$ = 9 $\rightarrow$ 8 line is always accompanied with a CO $J$ = 8 $\rightarrow$ 7 detection; there is one location (18$^h$:53$^m$18.5$^s$, 1$^\circ$:27':10'') in the F2 map where the CO $J$ = 9 $\rightarrow$ 8 integrated intensity is larger than the CO $J$ = 8 $\rightarrow$ 7 integrated intensity, due to the 9 $\rightarrow$ 8 line having a larger FWHM. At this location, the CO $J$ = 8 $\rightarrow$ 7 line still has a larger peak intensity. There are four locations in the C1 map where the CO $J$ = 9 $\rightarrow$ 8 line is detected but the CO $J$ = 8 $\rightarrow$ 7 line is not, despite reasonably similar RMS values between the two lines. For three of these locations, the CO $J$ = 9 $\rightarrow$ 8 line is only tentatively detected. The exception is the pixel centered at (18$^h$:42$^m$:47.3$^s$, -4$^\circ$:04':21''), where the CO $J$ = 9 $\rightarrow$ 8 line has a strong detection. Given the generally low signal to noise of all of the detected lines, it is possible that these unusual line ratios are just due to noise, although the CO $J$ = 9 $\rightarrow$ 8 line may indeed be stronger than the CO $J$ = 8 $\rightarrow$ 7 line towards some, or all, of these locations, in particular the C1 location with a strong 9 $\rightarrow$ 8 detection. 

The average linewidth of detections ranges between 4 and 7 km s$^{-1}$, depending upon the line and the clump. There are no obvious, significant trends in the distribution of FWHM or central velocity in the four clumps, although the determination of these line properties is significantly limited by the low signal to noise of the data. Maps of these quantities are therefore not shown. No lines other than the CO lines are detected in these {\it Herschel} observations. 

In the C1, F1, and F2 maps, the central velocities of the detected lines are all within 3.5 km s$^{-1}$ of the expected velocity, with the most discrepant velocities typically being associated with either weaker detections or broader lines. For the one pixel from the G2 map where the CO $J$ = 8 $\rightarrow$ 7 line is detected and the spatially averaged CO $J$ = 8 $\rightarrow$ 7 spectrum of G2, however, the central velocities of the detected lines vary from the known velocity of the G2 clump by over 15 km s$^{-1}$. As such, we consider the G2 detections to be false positives, possibly caused by random noise in the spectra. The velocity discrepancy between the line detected in the G2 spatial average and the known G2 central velocity is unlikely to be due to an outflow, as protostellar outflows are typically well collimated, such that the outflow emission would only contribute to a few pointings within the surveyed area. In such a case, a detection in the G2 spatial average would require detections in multiple pixels, since map positions off of the outflow axis would contribute no signal to the spatial average. We also consider it highly unlikely that this detection is a line from a more complex molecular species, given that CO should be by far the most abundant molecule in these IRDCs. If the G2 detections are false positives due to noise, it is, however, unusual that the G2 CO $J$ = 9 $\rightarrow$ 8 spatially averaged WBS and HRS spectra show an emission feature at a level of three times the integrated intensity uncertainty, just below our threshold for a detection, at a similar 60 km s$^{-1}$ central velocity to the CO $J$ = 8 $\rightarrow$ 7 tentative detections, as the central velocities of noise induced false positives in different data sets should not necessarily be correlated.

\section{Comparison PDR model}
\label{PDR}

\subsection{Setup}
\label{PDR setup}

The \citet{Kaufman99} PDR model calculates the emission from a semi-infinite, uniform density slab of material that is illuminated from one side. The PDR model includes molecular freeze out and starts with roughly Solar chemical abundances. 

A microturbulent Doppler broadening of $\beta = 1.5$ km s$^{-1}$ is used, which corresponds to a FWHM of approximately 2.5 km s$^{-1}$, since FWHM $= 2 \beta \sqrt{\ln(2)}$. This is smaller than the FWHM observed for the mid-$J$ lines, but consistent with the linewidths of the previously observed $^{12}$CO $J$ = 3 $\rightarrow$ 2 line towards these clumps (\citealt{Sanhueza10}; Pon et al.\, in prep., Paper II). Single dish observations of the N$_2$D$^+$ $J$ = 2 $\rightarrow$ 1 line, which have a similar beam size to that of our {\it Herschel} observations, also reveal linewidths of the order of 2 km s$^{-1}$ \citep{Fontani11}. Using a smaller FWHM for the PDR model will increase the emission in the mid-$J$ lines, as the lower lines will be less effective coolants. 

To give the PDR models the best chance at matching the bright emission detected in the mid-$J$ CO lines, we use the larger PAH abundance introduced by \citet{Hollenbach12}, as this enhanced PAH abundance produces extra heating in the periphery of a cloud which leads to much larger mid-$J$ CO fluxes. 

The six cores surveyed have densities of the order of 10$^5$ cm$^{-3}$ \citep{Butler09, Butler12, Tan13,Butler14}, but the gas surrounding these cores is likely at a lower density closer to 10$^{4}$ cm$^{-3}$, as typically found in IRDCs \citep{Rathborne06, Tan14}. Most of the higher $J$ CO emission is also expected to come from the warmer, outer layers of a molecular cloud, where the density is lower than in the centers of dense cores. We thus compare our {\it Herschel} intensities to PDR models with densities of either 10$^4$ or 10$^5$ cm$^{-3}$. 

We examine models with interstellar radiation fields (ISRFs) of log($G_0$) = 0 ($G_0$ = 1 Habing), where the average far-ultraviolet ISRF in free space is 1.7 Habing, or $1.6 \times 10^{-3}$ erg cm$^{-2}$ s$^{-1}$. We also examine a model with a density of 10$^{4}$ cm$^{-3}$ and an ISRF of log($G_0$) = 0.5 (G$_0 \approx 3$ Habing), to test the effect of increasing the UV field. 

In all of the PDR models, the CO $J$ = 8 $\rightarrow$ 7 and higher lines are optically thin ($\tau < 0.5$), such that we double the integrated intensities of the lines to account for emission directed inwards from the far side of the cloud. Since the models go to an A$_V$ of 10.4 mag, this doubling accounts for the emission from a slab with an A$_V$ of 20.8 mag. While the visual extinctions towards the IRDC core centers are much higher than this, of the order of 100 mag, we consider the PDR model predictions to still be reasonably accurate, as most of the mid-$J$ CO emission comes from the warm outer layers where CO is not heavily frozen out. In the three models, over 90\% of the emission in the CO $J$ = 8 $\rightarrow$ 7, 9 $\rightarrow$ 8, and 10 $\rightarrow$ 9 lines is emitted from gas at A$_V \le 1.5$. 

\subsection{PDR results}
\label{PDR results}

\begin{figure*}
   \centering
   \includegraphics[width=6.5in]{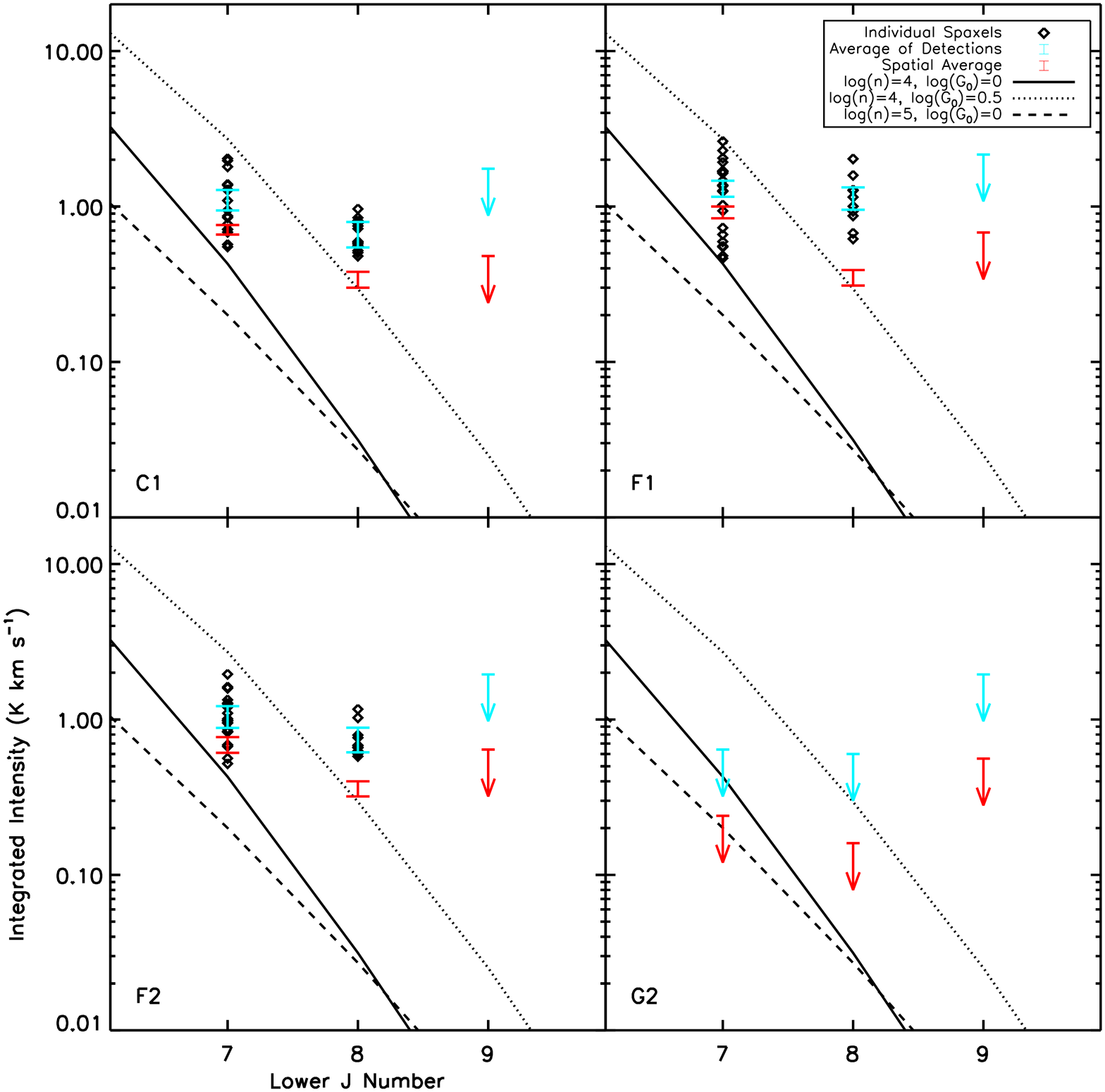}
   \caption{Integrated intensities of various CO rotational transitions as predicted by the \citet{Kaufman99} PDR models (lines). The solid line is for a PDR model with log($n$) = 4 and log($G_0$) = 0, the dotted line is for log($n$) = 4 and log($G_0$) = 0.5, and the dashed line is for log($n$) = 5 and log($G_0$) = 0. The black diamonds give the integrated intensities from individual pixels. The light blue points are centered on the mean value of all of the detections towards a particular clump and the error bars on these points are sized to the average uncertainty of all single pixel detections, such that the uncertainty in the average integrated intensity detection is much smaller. For maps with no detections, the light blue upper limit instead shows the average upper limit for an individual pixel. The red points show the integrated intensity of the average of all of the spectra across each map, with the error bars showing four times the uncertainty of this integrated intensity. The {\it top left, top right, bottom left}, and {\it bottom right} panels show the data for the C1, F1, F2, and G2 regions, respectively. Note how none of the PDR models can account for the large observed CO $J$ = 9 $\rightarrow$ 8 integrated intensities.}
   \label{fig:PDR}
\end{figure*}

Figure \ref{fig:PDR} shows the PDR model predictions as well as the observed integrated intensities of the mid-$J$ CO lines. All three PDR models underpredict the observed CO $J$ = 9 $\rightarrow$ 8 integrated intensities in the C1, F1, and F2 clumps, including the integrated intensities from the spatial averages, which have almost an order of magnitude lower integrated intensity than the maximum CO $J$ = 9 $\rightarrow$ 8 integrated intensity measured from an individual pixel. The models with an ISRF of 1 Habing underpredict the observed CO $J$ = 8 $\rightarrow$ 7 integrated intensity, although the higher ISRF model for a density of 10$^4$ cm$^{-3}$ is capable of producing an 8 $\rightarrow$ 7 integrated intensity larger than the largest observed integrated intensity of the 8 $\rightarrow$ 7 line. The non-detection of the CO $J$ = 10 $\rightarrow$ 9 line in all of the spectra is consistent with all three PDR models, while the non-detection of any mid-$J$ CO lines in the G2 clump is only consistent with the highest density PDR model. The 0.26 K km s$^{-1}$ upper limit on the CO $J$ = 10 $\rightarrow$ 9 integrated intensity from the combined spectrum of all four clumps is still consistent with all three PDR models.
 
The log($n$) = 4, log($G_0$) = 0 model produces a large CO $J$ = 8 $\rightarrow$ 7 to 9 $\rightarrow$ 8 integrated intensity ratio of 13.6, while the higher ISRF model for this density produces a slightly lower ratio of 9.2. The log($n$) = 5, log($G_0$) = 0 model generates a ratio of 7.4. For all three models, the ratio of the CO $J$ = 8 $\rightarrow$ 7 to 9 $\rightarrow$ 8 integrated intensities is nowhere near the average observed ratios of 1.6 to 2.0 for individual pixels in the C1, F1, and F2 maps. The models are also highly inconsistent with the CO $J$ = 8 $\rightarrow$ 7 to 9 $\rightarrow$ 8 integrated intensity ratios obtained from the spatially averaged spectra, which are in the range of 1.9 to 2.6 for the C1, F1, and F2 maps. 

\section{DISCUSSION}
\label{discussion}

	Based on the PDR models, the CO $J$ = 9 $\rightarrow$ 8 line should not have been detected in any pixel or in any of the spatially averaged spectra. Instead, the observations clearly show CO $J$ = 9 $\rightarrow$ 8 detections in over 25\% of the pixels in the C1, F1, and F2 maps, as well as in the spatially averaged spectra of these maps. Similarly, the PDR models predict a much larger CO $J$ = 8 $\rightarrow$ 7 to 9 $\rightarrow$ 8 integrated intensity ratio than observed, with the discrepancy between the highest observed ratio calculated from the spatially averaged spectra and the lowest predicted ratio still being 12 times the uncertainty for the observed ratio. While the average ratio from pixels with detections in both lines may be slightly skewed to lower ratios, since requiring both lines to be detected preferentially selects pixels with larger CO $J$ = 9 $\rightarrow$ 8 integrated intensities, the spatially averaged spectra, which do not have this selection bias, produce similarly low ratios for the C1, F1, and F2 clumps.

	Increasing the density of the PDR does not significantly change the absolute integrated intensities of the mid-$J$ lines, but rather primarily decreases the ratio of the CO $J$ = 8 $\rightarrow$ 7 to 9 $\rightarrow$ 8 integrated intensities. Raising the ISRF increases the intensities of all transitions and slightly lowers the CO $J$ = 8 $\rightarrow$ 7 to 9 $\rightarrow$ 8 ratio. For relatively weak ISRFs ($G_0$ $<$ 100), increasing the ISRF has a very limited effectiveness in creating hot CO, as larger ISRFs increase the size of the bounding C$^+$ layer, causing CO to form at larger extinctions, where the heating rate, and thus temperature, are lower. CO is also such an effective coolant, that upon formation of significant quantities of CO, the gas temperature tends to drop rapidly. This suggests that a PDR region with a much higher density and ISRF would be needed to consistently match both the absolute integrated intensities of the CO $J$ = 8 $\rightarrow$ 7 and 9 $\rightarrow$ 8 lines and the ratio between these lines. Given the observational constraints on the density and ISRF of these regions (see Sect.\ \ref{sources}), it seems highly unlikely that PDR emission alone can explain the observed mid-$J$ CO line intensities. The inability of the PDR models to reproduce the large observed CO $J$ = 9 $\rightarrow$ 8 integrated intensities and the CO $J$ = 8 $\rightarrow$ 7 to 9 $\rightarrow$ 8 integrated intensity ratios strongly suggests that there is a gas component in the C1, F1, and F2 clumps that is not accounted for by the PDR models.

	RADEX \citep{Vandertak07} 1D, non-local thermodynamic equilibrium (LTE) radiative transfer models with CO column densities of the order of $2 \times 10^{18}$ cm$^{-2}$, line widths of 5 km s$^{-1}$, densities of 10$^5$ cm$^{-3}$, and gas temperatures around 20 K also produce CO $J$ = 8 $\rightarrow$ 7 to 9 $\rightarrow$ 8 intensity ratios of the order of 10. Only at temperatures above 40 K does the line ratio approach the observed values, although such high temperature models, with the full CO column density of $2 \times 10^{18}$ cm$^{-2}$, significantly overproduce, by an order of magnitude, the observed integrated intensities. Radex models with lower column densities, such that the CO $J$ = 8 $\rightarrow$ 7 and 9 $\rightarrow$ 8 lines are optically thin, suggest a ratio between the lines of approximately 2 is only reached at temperatures above 100 K. Thus, the low observed ratio of the CO $J$ = 8 $\rightarrow$ 7 to 9 $\rightarrow$ 8 lines suggests that the missing gas component in the PDR models is a hot gas component, with a temperature of at least 40 K, and likely above 100 K.
	
The variation of the mid-$J$ CO line strengths and ratios across the surveyed regions implies that the generating source of the hot gas component must be spatially intermittent and cannot be a relatively uniform heating source, such as cosmic ray heating. The beam size changes by less than 10\% between the CO $J$ = 8 $\rightarrow$ 7 and 9 $\rightarrow$ 8 transitions, such that differing beam filling factors due to changes in the beam size cannot account for the line ratio differences between the observations and the models.

The dissipation of turbulent motions in low velocity shocks is expected to produce a hot gas component filling a small volume fraction of the cloud, and such a hot gas component has already been observed towards a low mass star-forming region, Perseus B1-E5 \citep{Pon12Kaufman, Pon14Kaufman}. The large mid-$J$ CO integrated intensities towards the C1, F1, and F2 regions may be due to such a low volume filling factor hot gas component generated by the dissipation of turbulence. Preliminary models of turbulence dissipating in low velocity shocks propagating in 10$^5$ cm$^{-3}$ gas predict CO $J$ = 8 $\rightarrow$ 7 and 9 $\rightarrow$ 8 integrated intensities on the order of 1.5  and 1.0 K km s$^{-1}$, respectively, which are comparable to the observed intensities in the C1, F1, and F2 clumps (Pon et al., in prep., Paper III). 

While the C1-N, C1-S, F1, and F2 cores are believed to be starless, there are more evolved sources within IRDCs C and F that could be responsible for generating a warm gas component. For instance, as shown in the middle panel of Fig.\ \ref{fig:sigmaradec}, there is a 24 micron source towards the northeast corner of the F1 clump \citep{Shepherd07, Chambers09}, indicative of an embedded young stellar object. \citet{Foster14} have also identified an embedded population of low mass protostars near this 24 micron source that could heat gas from protostellar outflows. This 24 micron source location is roughly coincident with the largest CO $J$ = 8 $\rightarrow$ 7 and 9 $\rightarrow$ 8 integrated intensities detected in any map. The mid-$J$ CO emission in the F2 map also preferentially comes from the vicinity of the MM7 clump, which is known to contain a water maser, a signpost of embedded high mass star formation \citep{Chambers09}. It is less obvious, however, what embedded sources, if any, could be responsible for the mid-$J$ emission in the C1 map. The average FWHM of the mid-$J$ CO lines is of the order of 5 km s$^{-1}$, which is slightly larger than the 3 km s$^{-1}$ line width of the $^{12}$CO $J$ = 3 $\rightarrow$ 2 line seen towards these clumps (\citealt{Sanhueza10}; Paper II). The larger linewidths of the higher lines could be indicative of broadening from outflows, but could also be due to the low signal to noise of the mid-$J$ detections. High velocity outflows generate very strong shocks that can lead to gas temperatures above 1000 kelvin (e.g., \citealt{Kaufman96II}), which would create different line ratios compared to lower velocity shocks.

While the ratios of the mid-$J$ CO lines can discriminate between different origins for the hot gas component by constraining the temperature of the hot gas, Fig. \ref{fig:PDR} shows that the CO $J$ = 8 $\rightarrow$ 7 emission could come from either the hot gas component or the ambient gas, depending upon the density of the gas and the ISRF. Observations of lower $J$ CO lines are thus vital for constraining the PDR parameters and determining what fraction of the 8 $\rightarrow$ 7 emission comes from the hot gas component. Observations of lower $J$ CO transitions towards the C1, F1, and F2 clumps will be presented in a future paper (Paper II). The similarity in the spatial distribution of the CO $J$ = 8 $\rightarrow$ 7 and 9 $\rightarrow$ 8 emission, however, suggests that the two lines are powered by the same excitation process, such that the 8 $\rightarrow$ 7 emission may also come primarily from the warm gas component. 

	The G2 region is clearly different from the C1, F1, and F2 regions, in that no emission in any of the observed lines is detected in the G2 maps. It is not obvious, however, whether there are significant differences between the C1, F1, and F2 regions, given the large variance in emission between pixels in each of the individual maps. The lack of detections in the individual pixels in G2 can be attributed to the lack of a hot gas component within G2, or at least a less substantial hot gas component in G2 than in the other three observed regions. Of the four clumps observed, the G2 clump has the smallest measured N$_2$H$^+$ line widths \citep{Fontani11}, such that G2 appears to be less turbulent than the other clumps. This lower level of turbulence may account for a reduced hot gas component, if the hot gas components in the C1, F1, and F2 clumps are due to turbulent dissipation. The lower turbulent state of G2 may be related to its evolved state, given that a transition to coherence is seen in move evolved, lower mass prestellar cores \citep{Pineda10}. The G2 clump also has almost twice the D/H ratio of the other three clumps as measured from the ratio of N$_2$D$^+$ and N$_2$H$^+$ \citep{Fontani11} and an order of magnitude larger D/H ratio measured from NH$_3$ and NH$_2$D \citep{Fontani14}, thus suggesting that the G2 clump is cooler and potentially more evolved. Alternatively, if G2 is younger, then G2 would be more likely to be free of embedded low mass protostars and the associated heating from these protostars and their protostellar outflows. 
	
	While the non-detection of the CO $J$ = 8 $\rightarrow$ 7 line in the spatial average of the spectra from the G2 map is consistent with the highest density PDR model, the G2 clump is believed to have the lowest density of the four clumps observed \citep{Butler09, Butler12}. A lower PAH abundance than assumed in the PDR models would lead to cooler temperatures at low visual extinctions, which would in turn reduce the expected integrated intensities of the mid-$J$ CO lines from the PDR zone. As such, a lower PAH abundance in the G2 clump would allow the G2 clump non-detection to be still consistent with the expected CO $J$ = 8 $\rightarrow$ 7 emission from a lower density PDR zone around the G2 clump. 
		
\section{CONCLUSIONS}
\label{conclusions}

Maps of the CO $J$ = 8 $\rightarrow$ 7, 9 $\rightarrow$ 8, and 10 $\rightarrow$ 9 transitions were obtained with the {\it Herschel} Space Observatory towards the C1, F1, F2, and G2 clumps. These four clumps are all embedded within IRDCs and each contains at least one core that appears to be quiescent and prestellar. The CO $J$ = 8 $\rightarrow$ 7 and 9 $\rightarrow$ 8 lines are detected in roughly 25 to 50\% of the pixels towards C1, F1, and F2, but are not reliably detected in any pixel or in the spatial average of all pixels towards the G2 clump. The CO $J$ = 10 $\rightarrow$ 9 line is not detected in any pixel or any spatial average, thereby placing a 0.5 K km s$^{-1}$ upper limit on the integrated intensity in the spatially averaged spectra. The detected CO $J$ = 8 $\rightarrow$ 7 and 9 $\rightarrow$ 8 lines have typical integrated intensities of the order of 1.2 and 0.9 K km s$^{-1}$ in individual pixels and integrated intensities of the order of 0.8 and 0.35 K km s$^{-1}$ in the spatially averaged spectra. The emission is, however, not spatially uniform, with the line emission varying by at least a factor of four between different regions within the same map. The average integrated intensity ratio of the CO $J$ = 8 $\rightarrow$ 7 to 9 $\rightarrow$ 8 lines is found to be of the order of 1.6 to 2.6

PDR models based on \citet{Kaufman99} were compared to the observed integrated intensities. While the CO $J$ = 8 $\rightarrow$ 7 integrated intensities can be explained by PDR models with reasonable densities and ISRFs, the models all underpredict the CO $J$ = 9 $\rightarrow$ 8 integrated intensity, with the models predicting that the CO $J$ = 9 $\rightarrow$ 8 line should have been undetected at all locations given the RMS level obtained. The PDR models also produce CO $J$ = 8 $\rightarrow$ 7 to 9 $\rightarrow$ 8 integrated intensity ratios between 7 and 14, which are significantly larger than the observed ratios. The combination of abnormally high integrated intensities of the CO $J$ = 9 $\rightarrow$ 8 line and low ratios between the CO $J$ = 8 $\rightarrow$ 7 and 9 $\rightarrow$ 8 lines strongly suggests the presence of a hot gas component within the C1, F1, and F2 clumps with a temperature likely above 100 K. The presence of a hot gas component is consistent with predictions for the dissipation of turbulence in low velocity shocks \citep{Pon12Kaufman}, but could also be due to heating from outflow generated shocks or embedded protostars. The emission around the F1 and F2 clumps, in particular, is concentrated near regions with known star formation, although there are no obvious embedded sources associated with the emission regions near the C1 clump.

\begin{acknowledgements}
	We would like to thank our referee, Dr.\ Goldsmith, for helping us improve the quality of this paper. The authors would like to thank Dr.\ N.\ Bailey, Dr.\ J.\ Bailey, and Dr.\ J.\ D.\ Henshaw for many insightful conversations regarding the data presented in this paper. The authors also thank Dr.\ C.\ McCoey and Dr.\ S.\ Beaulieu for help with HIPE. A.\ P.\ and P.\ C.\ acknowledge the financial support of the European Research Council (ERC; project PALs 320620). D.\ J.\ acknowledges support from a Natural Sciences and Engineering Research Council (NSERC) Discovery Grant. I.\ J.\ S.\ acknowledges the funding received from the People Programme (Marie Curie Actions) of the European Union's Seventh Framework Programme (FP7/2007-2013) under REA grant agreement PIIF-GA-2011-301538. This research has made use of the Smithsonian Astrophysical Observatory (SAO) / National Aeronautics and Space Administration's (NASA's) Astrophysics Data System (ADS). HIFI has been designed and built by a consortium of institutes and university departments from across Europe, Canada and the United States under the leadership of SRON Netherlands Institute for Space Research, Groningen, The Netherlands and with major contributions from Germany, France and the US. Consortium members are: Canada: CSA, U.Waterloo; France: CESR, LAB, LERMA, IRAM; Germany: KOSMA, MPIfR, MPS; Ireland, NUI Maynooth; Italy: ASI, IFSI-INAF, Osservatorio Astrofisico di Arcetri-INAF; Netherlands: SRON, TUD; Poland: CAMK, CBK; Spain: Observatorio Astron\'{o}mico Nacional (IGN), Centro de Astrobiolog\'{i}a (CSIC-INTA). Sweden: Chalmers University of Technology - MC2, RSS \& GARD; Onsala Space Observatory; Swedish National Space Board, Stockholm University - Stockholm Observatory; Switzerland: ETH Zurich, FHNW; USA: Caltech, JPL, NHSC. This research has made use of the astro-ph archive. Some spectral line data were taken from the Spectral Line Atlas of Interstellar Molecules (SLAIM) (Available at http://www.splatalogue.net). (F. J. Lovas, private communication, \citealt{Remijan07}). 
\end{acknowledgements}

\bibliographystyle{aa}
\bibliography{ponbib}

\end{document}